\documentclass[12pt]{article}
\usepackage[utf8]{inputenc}
\usepackage[english]{babel}
\usepackage{babel}
\usepackage{a4wide}
\usepackage[colorlinks=false]{hyperref}
\usepackage{amsfonts}
\usepackage{amssymb}
\usepackage{graphics}
\usepackage{amsmath}
\usepackage{cite}
\usepackage{appendix}
\usepackage{etoolbox}
\usepackage{physics}
\usepackage{sectsty}
\usepackage{appendix}
\usepackage{float}
\usepackage{graphicx}
\usepackage{subfig}
\usepackage{systeme}
\usepackage{hyperref}
\usepackage{xcolor}
\usepackage[normalem]{ulem}
\usepackage[makeroom]{cancel}

\DeclareMathOperator{\SU}{SU(2)}
\DeclareMathOperator{\SL}{SL(2,R)}
\DeclareMathOperator{\su}{\mathfrak{su}(2)}
\DeclareMathOperator{\dif}{d\!}
\DeclareMathOperator{\e}{e}
\DeclareMathOperator{\im}{i}
\DeclareMathOperator{\K}{K}
\DeclareMathOperator{\F}{F}
\DeclareMathOperator{\E}{E}

\def\texto
       {\topmargin -20pt
        \oddsidemargin 0pt
        \headheight 0pt
        \headsep 0pt
        \textwidth 6.35in
        \textheight 9.25in
        }
\texto

\def\baselinestretch{1.2}
\sectionfont{\large}
\renewcommand{\theequation}{\thesection.\arabic{equation}}
\csname @addtoreset\endcsname{equation}{section}

\begin{document} 

\begin{titlepage}

\begin{center}
\hfill CERN-TH-2022-104

\phantom{\LARGE X}

\vskip 0.25in

{\large \textbf{Spinning strings: \texorpdfstring{$\boldsymbol{\lambda}$}{lambda}-deformation and non-Abelian T-dual limit}}

\vskip 0.45in

\textbf{Rafael Hern{\'a}ndez},\footnote{\href{mailto:rafael.hernandez@fis.ucm.es}{\texttt{rafael.hernandez@fis.ucm.es}}}$^{a}$ \textbf{Roberto Ruiz},
\footnote{\href{mailto:rafael.hernandez@fis.ucm.es}{\texttt{roruiz@ucm.es}}}$^{a}$ and  \textbf{Konstantinos Sfetsos} \footnote{\href{mailto:ksfetsos@phys.uoa.gr}{\texttt{ksfetsos@phys.uoa.gr}}}$^{bc}$

\setcounter{footnote}{0} 

\vskip 0.15in

$\vphantom{x}^{a}$\! Departamento de F{\'i}sica Te{\'o}rica \\
and \\
Instituto de F{\'i}sica de Part{\'i}culas y del Cosmos (IPARCOS), \\
Facultad de Ciencias F{\'i}sicas, \\
Universidad Complutense de Madrid, \\
28040 Madrid, Spain \\ 

\vskip 0.15in

$\vphantom{x}^{b}$\!
Department of Nuclear and Particle Physics, \\ 
Faculty of Physics, \\
National and Kapodistrian University of Athens, \\
15784 Athens, Greece \\

\vskip 0.15in

$\vphantom{x}^{c}$\! Theoretical Physics Department,
 CERN, 1211 Geneva 23, Switzerland

\vskip 0.1in 

\end{center}

\vskip .5in

\centerline{\bf Abstract} 

\vskip .1in

\begin{sloppypar}

\noindent
The simplest example of the $\lambda$-deformation connects the $\SU$ Wess-Zumino-Witten model with the non-Abelian T-dual (NATD) of the $\SU$ principal chiral model. 
We analyze spinning strings with one spin propagating through the truncation of an S-dual embedding of the $\lambda$-deformation of the target space into type IIB supergravity.
We show that the situation apart from the NATD limit parallels the undeformed case. We demonstrate that regular spinning strings are either folded or circular, and that nearly degenerate spinning strings are either nearly point-like, fast, or slow. 
The effects of the \mbox{$\lambda$-deformation} 
are both the overall increment of the energy of spinning strings and the enlargement of the gap between the energies of folded and circular strings. In the NATD limit, 
we prove that circular strings disappear and that fast strings realize the dispersion relation of \mbox{Gubser-Klebanov-Polyakov} strings.

\end{sloppypar}

\vskip .4in

\vskip .1in

\noindent

\end{titlepage} 

\def\baselinestretch{1.2}
\baselineskip 20pt
\sectionfont{\large} 
\renewcommand{\theequation}{\thesection.\arabic{equation}} 
\csname @addtoreset\endcsname{equation}{section} 

\vfill
\eject

\tableofcontents


\section{Introduction} 
\label{int} 

The semiclassical limit is a central piece of our current understanding of quantum field theory. In the $\mathrm{AdS}/\mathrm{CFT}$ correspondence, the semiclassical limit involves classical strings as solutions 
to the equations of motion of the nonlinear \mbox{$\sigma$-model} on $\mathrm{AdS}_{5}\times\mathrm{S}^{5}$ that carry semiclassical quantum numbers~\cite{0202021,0204051,0204226,0304255}. 
The semiclassical spectrum organizes \mbox{non-supersymmetric} states on $\mathrm{AdS}_{5}\times\mathrm{S}^{5}$ and can be compared directly with the set of conformal dimensions 
of $\mathcal{N}=4$ supersymmetric Yang-Mills theory. 

Spinning strings on $\mathrm{AdS}_{5}\times\mathrm{S}^5$ are particularly significant classical solutions. They realize non-supersymmetric states with semiclassical and large total angular momentum~$J$ 
in the five-sphere~\cite{0304255}. The energy~$E$ of spinning strings admits a series expansion in~$J$, in parallel with the energy $E$ of non-supersymmetric Gubser-Klebanov-Polyakov (GKP) strings, which admits 
a series expansion in the Lorentzian spin~\cite{0204051}. Furthermore, the series expansion of the energy~$E$ of spinning strings can be matched with conformal dimensions in $\mathcal{N}=4$ supersymmetric Yang-Mills theory, 
just as the energy $E$ of supersymmetric \mbox{Berestein-Maldacena-Nastase} (BMN) point particles. This fact triggered intense research, where the systematization by the spinning-string ansatz stood out \cite{0307191,0311004}. 
We refer to \cite{0311139,1012.3986} for reviews of classical strings on $\mathrm{AdS}_{5}\times\mathrm{S}^5$ and a more complete set of references. 

The importance of classical strings on $\mathrm{AdS}_{5}\times\mathrm{S}^{5}$ raises the question of whether they can shed light on the semiclassical spectrum of general nonlinear $\sigma$-models. 
A renowned class of models is provided by the $\lambda$-deformation. The $\lambda$-deformation, introduced in \cite{1312.4560}, generalizes the current-current deformation of models 
with left and right current symmetry algebras. The key property of the $\lambda$-deformation is that it systematically yields the Lagrangian that accounts for all-loop perturbative effects 
in the deformation parameter~$\lambda$. Other notable properties are the general breaking of target-space isometries and the preservation of integrability in certain cases~\cite{1312.4560,1407.2840,1409.1538,1412.5181,1612.05012,1704.07834,
1707.05149,1806.10712,1809.03522,1812.04033,1902.04142,1909.02618,1911.07859,2005.02414,2106.00032,2012.08527,2202.08535}.  
The $\lambda$-deformation of some target spaces can be promoted 
to supergravity backgrounds~\cite{1410.1886,1504.02781,1504.07213,1601.08192,1606.00394,1608.03570,1911.12371,2103.12761}.  
The simplest example is the isotropic $\lambda$-deformation of the Wess-Zumino-Witten (WZW) model at level~$k$ based on the Lie group~$\mathrm{G}$~\cite{1312.4560}. 
The $\lambda$-deformation of the WZW model follows from four steps. First, the introduction of an auxiliary principal chiral model (PCM) based on $\mathrm{G}$. Second, 
the gauging of~$\mathrm{G}_{L}\times\widetilde{\mathrm{G}}_{D}$, where $\mathrm{G}_{L}$ denotes the left symmetry subgroup of the PCM 
and~$\widetilde{\mathrm{G}}_{D}\subset\widetilde{\mathrm{G}}_{L}\times\widetilde{{\mathrm{G}}}_{R}$ denotes the diagonal subgroup of the 
current symmetry group of the WZW model. Third, the imposition of the gauge-fixing condition whereby the PCM decouples. Finally, the integration of the gauge fields. 
The upshot is an effective action that, at leading order in~$1/k$, accounts for the \mbox{all-loop} effects of the $\lambda$-deformation. The deformation parameter~$\lambda$ 
is unique if the $\lambda$-deformation is isotropic, in which case its range is~$0\leqslant\lambda\leqslant 1$. (We will restrict ourselves to isotropic $\lambda$-deformations, 
and, hence, we will not make any further comment on isotropy in the following.) The effective action is invariant under the right symmetry subgroup $\mathrm{G}_{R}$ of the PCM. 
The $\lambda$-deformation interpolates between two models. On the one hand, the effective action reproduces the usual current-current deformation of the WZW model if~$\lambda\rightarrow0$. 
On the other hand, the effective action realizes the \mbox{Non-Abelian} T-dual (NATD) of the PCM with respect to~$\mathrm{G}_{L}$ if ~$\lambda=1$. The NATD limit involves both~$\lambda\rightarrow1$, 
which corresponds to~$k\rightarrow\infty$, and the rescaling of the coordinates of~$\mathrm{G}$. The NATD limit appears in the guise of a zoom-in limit, which justifies the decompactification implied 
by the NATD limit and solves associated global ambiguities. The $\lambda$-deformation that interpolates between the gauged WZW model on the coset~$\mathrm{G}/\mathrm{H}$ and the NATD 
of the coset model on~$\mathrm{G}/\mathrm{H}$ follows from analogous steps to the previous ones~\cite{1312.4560}. 
Furthermore, the $\lambda$-deformation can be adapted to symmetric spaces~\cite{1407.2840} and semi-symmetric spaces~\cite{1409.1538}. Despite similarities, the $\lambda$-deformation does not preserve any symmetry subgroup in these cases. 

The $\lambda$-deformation offers a unique panorama of seemingly different nonlinear $\sigma$-models. Classical strings that propagate through the target space of the $\lambda$-deformation 
of nonlinear $\sigma$-models should clarify the semiclassical limit of their superstring embedding if present. Based on integrability,~\cite{1704.05437} first addressed this problem by analyzing giant magnons 
on the \mbox{$\lambda$-deformation} of $\mathrm{AdS}_{5}\times\mathrm{S}^{5}$. Giant magnons follow from the application of the dressing method to the vacuum. The quantum-deformed 
isometry superalgebra of the S-matrix preserved by the vacuum satisfies a shortening condition that implies the dispersion relation of giant magnons. This result matches exactly the semiclassical 
dispersion relation that follows from the monodromy matrix of the Lax connection. One remarkable property of the giant magnons of \cite{1704.05437} is that their conserved charges have no known realization 
in the target space since the $\lambda$-deformation completely breaks the isometry supergroup of $\mathrm{AdS}_{5}\times\mathrm{S}^{5}$. Moreover, \cite{2111.12446} addressed the problem 
of the $\lambda$-deformation of classical strings from another point of view. This work considered the $\lambda$-deformation of~$\SU/\mathrm{U}(1)$ and~$\SL/\mathrm{U}(1)$, which are embeddable in supergravity \cite{1410.1886}, and used that the Virasoro constraints 
of a coset model imply its equations of motion if the target space is \mbox{two-dimensional}. This  relationship permitted to construct broad classes of closed and open classical strings. In the approach of \cite{2111.12446}, 
it is an open question how to obtain information about the spectrum. 

In this article, we will consider spinning strings that propagate through the target space of the \mbox{$\lambda$-deformation} of the~$\SU$ WZW model~\cite{1312.4560}.
The target space of the interpolation is~$\mathrm{S}^{3}_{\lambda}$, the \mbox{$\lambda$-deformation} of~$\mathrm{S}^3$ with pure Neveu-Schwarz-Neveu-Schwarz (NSNS) flux. We will assume that the target space is embeddable into a type IIB supergravity background with a number of properties. 
To construct spinning strings, we will adopt an alternative approach to~\cite{1704.05437,2111.12446}. We will exploit that the \mbox{$\lambda$-deformation} preserves
the symmetry subgroup~$\mathrm{SU}(2)_{R}$ of the~$\SU$ WZW model and the \mbox{S-duality} of type IIB superstring theory. We must mention that we are able to construct spinning strings despite the fact we should expect the general violation of classical integrability of the $\lambda$-deformation of the~$\SU$ WZW model by S-duality. These properties will permit us the use of the spinning-string ansatz of \cite{0307191}, which will reduce the equations of motion to a first-order ordinary differential equation (ODE). The ansatz will also allow for the computation of the dispersion relation of spinning strings. 

The organization of the paper is the following. In section \ref{sint}, we will review the \mbox{$\lambda$-deformation} of the~$\SU$ WZW model and introduce the spinning-string ansatz in the regime \mbox{$0\leqslant\lambda<1$}. 
In section~\ref{sstrings}, we will analyze spinning strings with $0\leqslant\lambda<1$. We will discriminate between regular spinning strings, and degenerate and nearly degenerate spinning strings. In section~\ref{sintNATD}, 
we will switch to the NATD limit, where $\lambda=1$. We will review the NATD limit of the \mbox{$\lambda$-deformation} of the $\SU$ WZW model and present the spinning-strings ansatz in this case. 
In section~\ref{sstringsNATD}, we will address spinning strings in the NATD limit. We will analyze spinning strings according to regularity classes again. In section \ref{sconclusions}, we conclude 
with some general remarks and prospects on future research. We will make extensive use of elliptic integrals throughout the text. We will follow the conventions of \cite{gradshteynryzhik}, some of whose formulae we will borrow. 

\section{The \texorpdfstring{$\boldsymbol{\lambda}$}{lambda}-deformation and the spinning-string ansatz}  
\label{sint} 

In this section, we will present the preliminary material to analyze the $\lambda$-deformation of spinning strings. In subsection \ref{ssaction}, we will write both the coordinate systems 
and the target-space fields of~$\mathrm{S}^{3}_{\lambda}$, our assumptions on the embedding background, and the classical action. In subsection~\ref{ssansatz}, we will introduce the spinning-string ansatz, which truncates the equations of motion to the ODE of a mechanical system.

\subsection{The classical action} 
\label{ssaction}

The $\lambda$-deformation of the $\SU$ WZW model is suitably written in hyperspherical angles~\cite{1312.4560}, namely, the angles $0\leqslant\theta\leqslant\pi$, $0\leqslant\beta\leqslant\pi$, 
and $0\leqslant\gamma<2\pi$ of $\mathrm{S}^{3}_{\lambda}$. The metric and \mbox{skew-symmetric} tensor fields of~$\mathrm{S}^{3}_{\lambda}$ read
\begin{equation}
\label{GS} 
\dif s^2=k\left[\frac{1+\lambda}{1-\lambda}\dif\theta^2+\frac{(1-\lambda^2)\sin^2\theta}{(1-\lambda)^2\cos^2\theta+(1+\lambda)^2\sin^2\theta}(\dif\beta^2 + \sin^2\beta\dif\gamma^2)\right]
\end{equation} 
and  
\begin{equation}
\label{BS}
B=k\left[-\theta+\frac{(1-\lambda)^2\cos\theta\sin\theta}{(1-\lambda)^2\cos^2\theta+(1+\lambda)^2\sin^2\theta}\right]\sin\beta\dif\beta\wedge\dif\gamma \ .
\end{equation}
The angles $\beta$ and $\gamma$ parametrize $\mathrm{S}^{2}$ inside $\mathrm{S}^{3}_{\lambda}$, and (\ref{GS}) and (\ref{BS}) are invariant under shifts of~$\gamma$.

The level of the WZW model is $k\in\mathbb{N}$ and the deformation parameter is
\begin{equation}
\label{lambda}
\lambda=\frac{k}{\kappa^2+k} \ ,
\end{equation}
where $\kappa$ denotes the coupling of the auxiliary $\SU$ PCM used in the construction of the $\lambda$-deformation of the $\SU$ WZW model. The range of~$\lambda$ is $0\leqslant\lambda\leqslant1$. 
The value $\lambda=0$ corresponds to the $\SU$ WZW model, where~$\kappa=\infty$ with~$k$ fixed, and, hence, the $\SU$ PCM decouples. The value $\lambda=1$ corresponds 
to the NATD limit of the $\SU$ PCM with respect to~$\SU_{L}$~\cite{1312.4560}. We will analyze the NATD limit in section \ref{sintNATD}, and restrict ourselves to $0\leqslant\lambda<1$ until then.  

The starting point of our analysis is the assumption that the~\mbox{$\lambda$-deformation} of the~$\SU$ WZW model is embeddable into a special type IIB supergravity background, 
where of the former is a consistent truncation.  The background must fulfill the following conditions.

First, the RR three-form flux of the background must have no components along $\mathrm{S}^{3}_{\lambda}$. 
This condition will enable the spinning-string ansatz (\ref{ansatz2}) below to consistently truncate the equations of motion.
Specifically, under this condition, we will assume the use of S-duality to swap the B-field (\ref{BS}) and the RR two-form potential. 
Since the resultant S-dual NSNS flux has no components along $\mathrm{S}^{3}_{\lambda}$, we will be able to ignore it. 
We stress that the use of S-duality implies that our analysis will apply to F1-strings in the S-dual background rather than the initial background. 
Second, the dilaton of the background must be constant. This assumption enables us to identify the components of the metric in the Einstein and string frames.
Third, to define the energy of spinning strings, we will need the existence of the time-like Killing vector $\partial/\partial t$ in the background. The associated direction must be topologically the real line
to avoid closed time-like curves in the target space, and, hence, we will choose~$-\infty<t<\infty$. We will also demand that $\partial/\partial t$ is globally defined for the energy to translate into a quantum number under quantization. 
Finally, the background must preserve the isometry group $\SU_{R}$ of the~\mbox{$\lambda$-deformation} of the $\SU$ WZW model, 
which permits to define the angular momentum, and the application of the NATD limit to the full background must yield a finite result.

We are now ready to write the classical action of closed bosonic strings that propagate through the truncation of the S-dual background to $\mathrm{S}_{\lambda}^{3}$ together with the \mbox{time-like} direction. The action is the conformally gauge-fixed Polyakov action
\begin{equation}
\label{S} 
\begin{split}
S&=-\frac{k}{4\pi}\int_{-\infty}^{\infty}\dif\tau\int_{0}^{2\pi}\dif\sigma\left(-\frac{1+\lambda}{1-\lambda}(-\dot{t}^2+t'^2)+\frac{1+\lambda}{1-\lambda}(-\dot{\theta}^2+\theta'^2)\right.\\
&\left.+\frac{(1-\lambda^2)\sin^2\theta}{(1-\lambda)^2\cos^2\theta+(1+\lambda)^2\sin^2\theta}[-\dot{\beta}^2+\beta'^2 + \sin^2\beta \,(-\dot{\gamma}^2+\gamma'^2)\,]\right)\ ,
\end{split} 
\end{equation} 
where $\tau$ and $\sigma$ parametrize the time-like and space-like worldsheet directions, and the dot and the prime denote derivatives with respect to $\tau$ and $\sigma$, respectively. 
The Wess-Zumino term is absent in the classical action because S-duality has exchanged the NSNS flux of~(\ref{BS}) with the RR three-form flux that, by assumption, has no components along~$\mathrm{S}_{\lambda}^{3}$. Moreover, we have looked at the time-like component of the metric in the background of~\cite{1410.1886} to fix the normalization of~$t$ with respect 
to~$\lambda$ in the classical action. This normalization will ease the application the NATD limit in subsection \ref{ssactionNATD}. Any other normalization consistent with the NATD limit implied by the embedding background would entail at most a rescaling of the kinetic term of $t$ (and of the energy of spinning strings as a consequence).~\footnote{Reference~\cite{1410.1886} embedded the $\lambda$-deformation of the $\SU$ WZW model into type IIB* supergravity. 
Apart from the problems that type IIB* supergravity poses \cite{9806146,9807127}, it is unclear how our spinning strings would organize the semiclassical spectrum in this case.  
The reason is that $t$ does not correspond to a globally defined time-like Killing vector in the background of \cite{1410.1886}, but just parametrizes an isometric direction in the truncation 
$\tilde{\alpha}=t$ and $\tilde{\beta}=\pi/2$ of (3.11) in \cite{1410.1886} (where $\tilde{\gamma}$ is not defined).} 

Moreover, we must supplement the classical action with two conditions: the Virasoro constraints  
\begin{equation}
\begin{split} 
\label{V1}
\frac{1+\lambda}{1-\lambda}(\dot{t}^2+t'^2)&=\frac{1+\lambda}{1-\lambda}(\dot{\theta}^2+\theta'^2) \\
&+\frac{(1-\lambda^2)\sin^2\theta}{(1-\lambda)^2\cos^2\theta+(1+\lambda)^2\sin^2\theta}[\,\dot{\beta}^2+\beta'^2 + \sin^2\beta (\dot{\gamma}^2+\gamma'^2)\,]
\end{split}
\end{equation} 
and
\begin{equation}
\label{V2}
\frac{1+\lambda}{1-\lambda}\,\dot{t}t'=\frac{1+\lambda}{1-\lambda}\,\dot{\theta}\theta'
+\frac{(1-\lambda^2)\sin^2\theta}{(1-\lambda)^2\cos^2\theta+(1+\lambda)^2\sin^2\theta}(\dot{\beta}\beta' + \sin^2\beta\,\dot{\gamma}\gamma'\,)\ , 
\end{equation}
and the periodic boundary conditions of the target-space coordinates
\begin{equation}
\label{bc}
\begin{split}
t(\tau,\sigma+2\pi)&=t(\tau,\sigma) \ , \quad \theta(\tau,\sigma+2\pi)=\theta(\tau,\sigma) \ , \\ \beta(\tau,\sigma+2\pi)&=\beta(\tau,\sigma) \ , \quad \gamma(\tau, \sigma+2\pi)=\gamma(\tau,\sigma) \ .
\end{split}
\end{equation}
Invariance of the classical action under shifts of $t$ and $\gamma$ implies the conservation of the energy~$E$ and the angular momentum~$J$, respectively, whose expressions follow from Noether's theorem. 
The angular momentum $J$ is associated to the generator of the Cartan subalgebra of~$\su_{R}$, which is the isometry algebra of the $\lambda$-deformation of the~$\SU$ WZW model. 
There are no two angular momenta since the $\lambda$-deformation breaks the isometry group~$\mathrm{SU}(2)_{L}\times\mathrm{SU}(2)_{R}$ at $\lambda=0$ (itself a subgroup of the full current isometry group) 
down to~$\mathrm{SU}(2)_{R}$.  

\subsection{The spinning-string ansatz} 
\label{ssansatz} 

The construction of spinning strings is based on the ansatz for the target-space coordinates presented in~\cite{0307191}. The ansatz consistently truncates the equations of motion, 
and the truncation follows from a mechanical system. In addition, the ansatz simplifies the expression of the conserved charges and facilitates the computation of the dispersion relation.

If $\lambda=0$, the spinning-string ansatz is defined through the embedding of $\mathrm{S}^{3}$ into $\mathrm{R}^{4}$~\cite{0307191}, 
\begin{equation}
\label{ansatz}
t=\mu\tau \ , \quad X^{1}+\im X^{2}=\e^{\im\omega_{1}\tau}\cos\psi (\sigma) \ , \quad X^{3}+\im X^{4}=\e^{\im\omega_{2}\tau}\sin\psi(\sigma) \ ,
\end{equation}
where $0\leqslant \psi(\sigma)\leqslant\pi/2$ is periodic. The ansatz (\ref{ansatz}) truncates the equations of motion of the $\SU$ WZW model to an ODE with respect to~$\sigma$, which, in turn, is the \mbox{Euler-Lagrange} equation 
of the Lagrangian of a mechanical system. The consistency of the truncation is a consequence of the invariance of the classical action under $\mathrm{SO}(2)$-rotations of the two planes 
at $X^{1}=X^{2}=0$ and~$X^{3}=X^{4}=0$, which ensures that~$\tau$ drops out of the equations of motion. This property is the specification of \mbox{$\mathrm{SU}(2)_{L}\times\mathrm{SU}(2)_{R}$}-rotations 
under which the classical action is invariant to shifts along the directions of the Cartan torus. 

If $\lambda>0$, the ansatz (\ref{ansatz}) no longer yields a consistent truncation. Since the embedding coordinates are related to the hyperspherical angles as
\begin{equation}
\label{embedding}
X^{1}=\cos\theta \ , \quad X^{2}=\sin\theta\cos\beta \ , \quad X^{3}=\sin\theta\sin\beta\cos\gamma \ , \quad X^{4}=\sin\theta\sin\beta\sin\gamma \ ,
\end{equation} 
we deduce from (\ref{GS}) and (\ref{BS}) that~(\ref{S}) is invariant under the $\mathrm{SO}(2)$-rotation of the plane at $X^{1}=X^{2}=0$, but not of the plane at $X^{3}=X^{4}=0$. 
This property is a consequence of the breaking of $\mathrm{SU}(2)_{L}\times\mathrm{SU}(2)_{R}$ down to $\mathrm{SU}(2)_{R}$ by the $\lambda$-deformation, which is also responsible 
for the conservation of one angular momentum instead of two. Therefore, the ansatz~(\ref{ansatz}) consistently truncates the equations of motion of~(\ref{S}) if and only if~$\omega_{1}=0$. If $\omega_{1}=0$, we deduce from (\ref{ansatz}) and (\ref{embedding}) that the ansatz that we must consider is~\footnote{We could generalize the spinning-string ansatz (\ref{ansatz2}) along the lines of \cite{0311004} by letting $\gamma=\omega\tau+\bar{\gamma}(\sigma)$.  The Virasoro constraint~(\ref{V2}) would fix $\bar{\gamma}=0$ nonetheless. In fact, we have not included winding numbers along the direction of $\gamma$ in the boundary conditions (\ref{bc}) for this reason.}
\begin{equation} 
\label{ansatz2}
t=\mu\tau \ , \quad \theta=\theta(\sigma) \ , \quad \beta=\frac{\pi}{2}  \ , \quad \gamma=\omega\tau \ ,
\end{equation}
where $\omega_{2}\equiv\omega$. We will assume $\omega\geqslant0$ without loss of generality. Due to (\ref{bc}), we must supplement~$\theta$ with periodic boundary conditions,
\begin{equation}
\label{bc2}
\theta(\sigma+2\pi)=\theta(\sigma) \ ,
\end{equation}
whereas we can safely  ignore the periodic boundary conditions of $t,$ $\beta$, and $\gamma$ since they trivially hold. The ansatz (\ref{ansatz2}) realizes the class of spinning strings with one spin, which \cite{0204051} analyzed on $\mathrm{AdS}_{5}\times\mathrm{S}^{5}$ for the first time. 

We should make a remark before moving forward. If we had introduced~(\ref{ansatz2}) in the equations of motion of the original $\lambda$-deformation of the $\SU$ WZW model, the contribution of the B-field (\ref{BS}) would have implied $\theta^{\prime}=0$. The ansatz~(\ref{ansatz2}) would have realized point particles as a consequence. It is now apparent that the type IIB supergravity  is crucial to drop the B-field by S-duality and allow (\ref{ansatz2}) to realize actual spinning strings.

We can already argue that the ansatz ({\ref{ansatz2}) describes rotating motion. First of all,~(\ref{ansatz2}) specifies the $\lambda$-deformation of the two-sphere $\mathrm{S}^{2}_{\lambda}$ inside~$\mathrm{S}^{3}_{\lambda}$. 
The pair~$\theta$ and~$\gamma$ are the polar and azimuthal angles of~$\mathrm{S}_{\lambda}^{2}$, respectively.~\footnote{We can see that $\theta$ and $\gamma$ parametrize the $\lambda$-deformation of a $\mathrm{S}^{2}$ if we set 
$\lambda=0$ and $\beta =\pi / 2$  in the metric (\ref{GS}). The result is $\dif s^2 =\dif\theta^2 + \sin^2\theta \dif\gamma^2$, which is the metric of $\mathrm{S}^2$ for $0\leqslant\theta\leqslant\pi$ and~\mbox{$0\leqslant\gamma<2\pi$}. 
The space $\mathrm{S}^{2}_{\lambda}$ should not be confused with the target space of the $\lambda$-deformation of the $\mathrm{SU}(2)/\mathrm{U}(1)$ coset model [9], which admits no Killing vectors.}
The angle $\gamma$ parameterizes, in particular, a great circle of $\mathrm{S}_{\lambda}^2$: a geodesic along the only isometric direction of $\mathrm{S}_{\lambda}^2$.
The ansatz $\gamma=\omega\tau$ states that strings are stretched along the meridians of~$\mathrm{S}_{\lambda}^{2}$. The ansatz~$\theta=\theta(\sigma)$ quantifies the extension of the strings along the meridians. 
Thus, strings rigidly rotate around the axis of~$\mathrm{S}_{\lambda}^{2}$ that crosses the poles at~$\theta=0$ and~$\theta=\pi$, and rotations are isochronous since $t=\mu\tau$. We have already mentioned that 
the ansatz~(\ref{ansatz2}) corresponds to point particles in particular if~\mbox{$\theta'=0$}. Point particles encompass BMN particles~\cite{0202021,0204051}.

If we introduce (\ref{ansatz2}) in the equations of motion of the classical action, we obtain an ODE that is the equation of motion of the Lagrangian of a mechanical system. 
This Lagrangian in fact  follows from the direct introduction of (\ref{ansatz2}) into the classical action.  Following~\cite{0307191}, we define the ellipsoidal coordinate $\zeta$, 
which is convenient because the Lagrangian is a rational function of~$\zeta$. The definition of ellipsoidal coordinate is
\begin{equation}
\label{zeta}
\cos^2\theta=\zeta\ , \quad \sin^2\theta=1-\zeta\ ,
\end{equation}
with $0\leqslant\zeta\leqslant1$. Note that $\cos\theta$ as a function of $\zeta$ has two branches: the positive branch with $0\leqslant\theta\leqslant\pi/2$ and the negative branch with $\pi/2\leqslant\theta\leqslant\pi$. 
The ansatz~(\ref{ansatz2}) corresponds to~$\zeta=\zeta(\sigma)$ with periodic boundary conditions.

The spinning-string ansatz eventually leads us to the Lagrangian 
\begin{equation} 
\label{L}
L = \frac{1}{2}\left[\frac{1+\lambda}{1-\lambda}\frac{\zeta'^2}{4\zeta(1-\zeta)}-\omega^2\frac{(1-\lambda^2)(1-\zeta)}{(1+\lambda)^2-4\lambda\zeta}\right] \ .
\end{equation}
The Virasoro constraints (\ref{V1}) and (\ref{V2}) force the classical trajectories of (\ref{L}) to realize spinning strings. The ansatz (\ref{ansatz2}) reduces the Virasoro constraint (\ref{V1}) to
\begin{equation}
\label{disp}
\frac{1+\lambda}{1-\lambda}\mu^2=H \ ,
\end{equation}
where $H$ is the Hamiltonian, corresponding to the mechanical energy of the system, 
\begin{equation}
\label{H}
H = \frac{1}{2}\left[\frac{1+\lambda}{1-\lambda}\frac{\zeta'^2}{4\zeta(1-\zeta)}+\omega^2\frac{(1-\lambda^2)(1-\zeta)}{(1+\lambda)^2-4\lambda\zeta}\right] \ .
\end{equation} 
On the other hand, (\ref{ansatz2}) trivializes the Virasoro constraint~(\ref{V2}).

The spinning-string ansatz simplifies the expressions of the conserved charges, namely, the energy $E$ and the angular momentum $J$  associated to the invariance of the classical action under shifts 
of $t$ and $\gamma$, respectively. In particular, (\ref{ansatz2}) reduces the energy to
\begin{equation}
\label{E0} 
E = - k\frac{1+\lambda}{1-\lambda}\,\mu \
\end{equation}
and the angular momentum to
\begin{equation}
\label{J0}
J=k\frac{1-\lambda^2}{2\pi}\int_{0}^{2\pi}\dif\sigma\, \frac{(1-\zeta)\omega}{(1+\lambda)^2-4\lambda\zeta} \ .
\end{equation}
Note that neither (\ref{E0}) nor (\ref{J0}) follow from the mechanical system, but they follow from the initial classical action instead. The Virasoro constraint (\ref{V1}) implies the proportionality 
between the energy of spinning strings~$E$ and the energy of the mechanical system~$H$.

A final remark is in order. Apart from the angular momentum~$J$, Noether's theorem implies, in general, the conservation of two extra conserved charges, which correspond to the remaining pair 
of generators of the isometry algebra~$\su_{R}$. However, the ansatz~(\ref{ansatz2}) implies that they vanish, and, thus, we can safely ignore them. 

\section{Spinning strings with \texorpdfstring{$\boldsymbol{0\leqslant\lambda<1}$}{0 lambda 1}}
\label{sstrings}

In this section, we analyze the $\lambda$-deformation of spinning strings. We will discriminate between regular spinning strings, and degenerate and nearly degenerate spinning strings. 
In subsection~\ref{ssregular}, we will consider the two regular classes of spinning strings, namely, folded and circular strings, and we will compute  the dispersion relation of each class by numerical means. In subsection~\ref{ssndeg}, we will address degenerate and nearly degenerate spinning strings, whose dispersion relation we will compute analytically. 
We will consider, in particular, point particles, nearly point-like strings, fast strings, and slow strings.

Our starting point is $H$ in (\ref{H}), which is a non-negative integral of motion. Being a first integral, $H$ permits us to recast the equation of motion of $\zeta$ as a first-order ODE
that is solvable by a direct integration. Introducing the ``moment of inertia"~$I$ by 
\begin{equation}
\label{mi}
H\equiv\frac{1}{2}\omega^2I \ , 
\end{equation}
the differential equation reads
\begin{equation}
\label{ode}
\begin{split}
\zeta ' =\pm \omega \sqrt{\frac{P(\zeta)}{\zeta_{2}-\zeta}} \ ,
\end{split}
\end{equation}
where we have defined the cubic polynomial
\begin{equation}
\label{P}
P(\zeta)\equiv C\zeta (1-\zeta)\,(\zeta-\zeta_{1}) \ ,
\end{equation}
and the parameters 
\begin{equation}
\begin{split}
\label{parameters}
C&\equiv4I_{1}\left(I_{2}- I\right) \ ,  \quad I_{1}\equiv \frac{1-\lambda}{1+\lambda} \ < \ I_{2}\equiv \frac{1-\lambda^2}{4\lambda} \ , \\
\zeta_{1}&\equiv \frac{(1+\lambda)^2}{4\lambda}\frac{I_{1}-I}{I_{2}-I} \ ,
 \quad \zeta_{2}\equiv \frac{(1+\lambda)^2}{4\lambda}\ > \ 1 \ .
\end{split}
\end{equation}
The $\pm$ signs in (\ref{ode}) correspond to the positive and negative branches of the square root, respectively. The classical trajectories that follow from (\ref{ode}) depend on~$\omega$ 
and~$I$ (together with~$\lambda$). Later, we will trade $\omega$ and $I$ for the angular momentum~$J$ and another quantum number (the mode number~$N$,  to be introduced below) in order to write the dispersion relation. 

We can draw some conclusions from (\ref{ode}). Classical trajectories must respect~$0\leqslant\zeta\leqslant1$, which ensures $0\leqslant\theta\leqslant\pi$. Since~$\zeta_{2}>1$, 
either~$\zeta_{1}\leqslant\zeta\leqslant1$ or $0\leqslant\zeta\leqslant1$ must hold. 
The endpoints of these intervals are the roots of~$P(\zeta)$, which are branch points of the square root of~(\ref{ode}) and turning points of $\zeta'$. The actual interval of~$\zeta$ 
for a given classical trajectory depends on~$I$, which determines both the sign of~$C$ and the value of $\zeta_{1}$.  

The pair of alternative ranges~$\zeta_{1}\leqslant\zeta\leqslant1$ and $0\leqslant\zeta\leqslant1$ reflect the existence of two regular classes of spinning strings. 
We will show in subsection \ref{ssregular} that the situation for~$0<\lambda<1$ is analogous to the situation for~$\lambda=0$~\cite{1311.5800}. 
Regular spinning strings are either folded or circular. Classical solutions in one class are not continuously connected with solutions in the other. We depict schematically the shape of folded and circular strings 
in figures \ref{foldedpicture} and \ref{circularpicture}, respectively.  Regular spinning strings also have nearly degenerate limits. In particular, nearly degenerate spinning strings correspond to the values 
of the angular momentum $J\rightarrow0$ and $J\rightarrow\infty$. We will prove in subsection \ref{ssndeg} that folded strings realize nearly point-like strings when~\mbox{$J\rightarrow0$} 
and fast strings when~\mbox{$J\rightarrow\infty$}, and that circular strings realize slow strings when~$J\rightarrow0$ and fast strings again when $J\rightarrow\infty$. 
Finally, the spinning-string ansatz gives rise to point particles if $\zeta^{\prime}=0$, which are either instantons or BMN particles. Instantons and folded strings are connected at~\mbox{$J=0$}. BMN particles play, in a sense, the r{\^o}le of threshold between folded and circular strings. 

We summarize the classification of spinning strings below. 

\begin{itemize} 
\item{Folded strings. The inequalities $0<I<I_{1}$ hold. The range of $\zeta$ is~$\zeta_{1}\leqslant\zeta\leqslant 1$ and the angular momentum $J$ satisfies $0<J<\infty$. 
\begin{itemize} 
\item{
Nearly point-like strings. Folded strings with $I\rightarrow 0$, where $\zeta_{1}\rightarrow1$ and $J\rightarrow0$. 
}
\item{
Fast strings. Folded strings with $I\rightarrow I_{1}^{-}$, where $\zeta_{1}\rightarrow0$ and $J\rightarrow\infty$. 
}
\end{itemize}
}
\item{Circular strings. The inequality $I>I_{1}$ holds. The range of $\zeta$ is~\mbox{$0\leqslant\zeta\leqslant1$} and the angular momentum $J$ satisfies $0<J<\infty$. 
\begin{itemize}
\item{
Slow strings. Circular strings with $I\rightarrow\infty$, where $\zeta_{1}\rightarrow\zeta_{2}$ and $J\rightarrow0$. 
}
\item{
Fast strings. Circular strings with $I\rightarrow I_{1}^{+}$ , where $\zeta_{1}\rightarrow0$ and $J\rightarrow\infty$. 
}
\end{itemize} 
}
\item{Point particles. The equality $\zeta^{\prime}=0$ holds identically.
\begin{itemize} 
\item{
Instantons. Point particles with $\zeta=0$, where $J=0$. 
}
\item{
BMN particles. Point particles with $\zeta=1$, where $0<J<\infty$. 
}
\end{itemize} 
}
\end{itemize} 

\begin{figure}[H]
\centering
\subfloat[]{\includegraphics[width=0.35\textwidth]{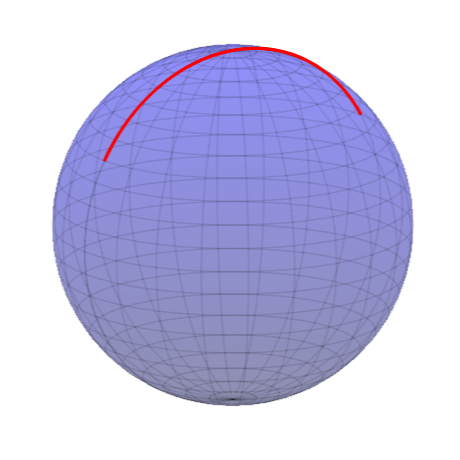}\label{foldedpicture}}
\subfloat[]{\includegraphics[width=0.35\textwidth]{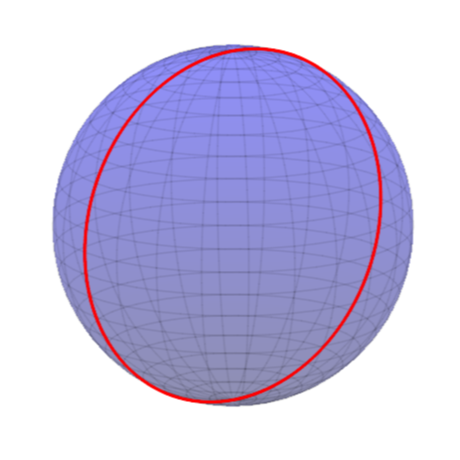}\label{circularpicture}}
\caption{Schematic depiction of the regular classes of spinning strings for $0\leqslant\lambda<1$ for fixed $\tau$. Figure \ref{foldedpicture} corresponds to folded strings and figure \ref{circularpicture} to circular strings. }
\end{figure}

\subsection{Regular spinning strings}  
\label{ssregular} 

In this subsection, we will analyze regular spinning strings. We will first present the solutions corresponding to folded and circular strings 
and we will conclude the analysis with the numerical computation of the dispersion relation of both classes. 

\subsubsection{Folded strings} 
\label{sssfolded} 

Let us analyze the class of spinning strings with $0< I< I_{1}$. First, we will solve (\ref{ode}). Since \mbox{$0< I< I_{1}$}, both $C>0$ and~$0<\zeta_{1}<1$ hold, and, 
thus, (\ref{ode}) implies $\zeta_{1}\leqslant\zeta\leqslant1$. If we use the invariance of~(\ref{ode}) under~$\sigma\mapsto\sigma+\sigma_{0}$ to fix $\zeta(0)=\zeta_{1}$, we must choose 
the positive branch of~(\ref{ode}). Let~$0\leqslant\sigma\leqslant\bar{\sigma}$ be the interval where $\zeta$ ranges between the two turning points of~$\zeta'$, namely~$\zeta(0)=\zeta_{1}$ and $\zeta(\bar{\sigma})=1$.  
The integration of (\ref{ode}) in this case leads us to
\begin{equation}
\begin{split}
\label{solutionI1} 
\omega\sigma &= \int_{\zeta_{1}}^{\zeta}\mathrm{d}x\sqrt{\frac{\zeta_{2}-x}{P(x)}}
=\left(\frac{1+\lambda}{1-\lambda}\right)^2\left[\F(\varphi,m)-\frac{4\lambda}{1-4\lambda I-\lambda^2}\left(\frac{1-\lambda}{1+\lambda}-I\right)\Pi(\varphi,n,m)\right] \ ,
\end{split} 
\end{equation} 
}
where 
\begin{equation}
\label{argumentsI}
\varphi=\sin^{-1}\sqrt{\frac{\zeta-\zeta_{1}}{(1-\zeta_{1})\,\zeta}} \ , \quad \ m=\sqrt{\frac{1+\lambda}{1-\lambda}I} \ , \quad \ n=\frac{(1-\lambda)^2I}{1-4\lambda I-\lambda^2} \ .
\end{equation}
The functions $\F(\varphi,m)$ and $\Pi(\varphi,n,m)$ are the incomplete elliptic integral of the first and third kind, respectively, and the branch of the inverse sine is $\sin^{-1}:\![0,1]\rightarrow[0,\pi/2]$. 
The conditions $0<m<1$ and $0<n<1$ guarantee that $\F(\varphi,m)$ and~$\Pi(\varphi,n,m)$ are analytic and real. We must stress that the inverse function of $\Pi(\varphi,n,m)$ does not admit a closed form, 
and, therefore, we cannot write a closed form for~$\zeta$ unless~$\lambda=0$. If $\lambda=0$,~(\ref{solutionI1}) just involves~$\F(\varphi,m)$, and we can write $\zeta$ in terms of the Jacobian elliptic sine. 

To extend (\ref{solutionI1}) along $\bar{\sigma}\leqslant\sigma\leqslant2\bar{\sigma}$, we must consider the negative branch of (\ref{ode}). In this interval,~$\zeta$ ranges between  
the turning points $\zeta(\bar{\sigma})=1$ and~$\zeta(2\bar{\sigma})=\zeta_{1}$. 
If we integrate~(\ref{ode}) now, we obtain
\begin{equation}  
\begin{split} 
\label{solutionI2}
\omega\sigma & =
\left( \int_{\zeta_{1}}^{1} -  \int_{1}^{\zeta}\right) \mathrm{d}x\sqrt{\frac{\zeta_{2}-x}{P(x)}}
=\left(\frac{1+\lambda}{1-\lambda}\right)^2\left[\,2\K(m)-\F(\varphi,m)
\vphantom{-\frac{4\lambda}{(1-\lambda)^2(1-4\lambda I-\lambda^2)}}\right.\\ &\phantom{x} \left.
-\frac{4\lambda}{1-4\lambda I-\lambda^2}\left(\frac{1-\lambda}{1+\lambda}-I\right)\big(2\Pi(n,m)-\Pi(\varphi,n,m)\big)\right] \ ,
\end{split}
\end{equation}
where $\K(m)$ is the complete elliptic function of the first kind and $\Pi(n,m)$ is the complete elliptic function of the third kind. As before, $\K(m)$ and $\Pi(n,m)$ are analytic 
and real since~$0<m<1$ and $0<n<1$. Note that (\ref{solutionI1}) and (\ref{solutionI2}) give the same expression at~\mbox{$\sigma=\bar \sigma$}, which is nothing but a reflection of the continuity 
of the classical solution. The extension of~(\ref{solutionI2}) to any other interval of~$\sigma$ straightforwardly follows by alternating of branches in~(\ref{ode}). 

We will now look at the shape of the string described by (\ref{solutionI1}) and (\ref{solutionI2}). First, we must note that the turning points of $\zeta'$ and $\theta'$ do not coincide. 
It follows from definition~(\ref{zeta}) that
\begin{equation} 
\label{zetaprime} 
\zeta'^2=4\sin^2\theta\cos^2\theta\,\theta'^2 \ , \qquad \theta'^2=\frac{\zeta'^2}{4\zeta(1-\zeta)} \ .
\end{equation} 
The ODE (\ref{ode}) implies that the turning points $\zeta=0$ and $\zeta=1$ of $\zeta^{\prime}$ are not turning points of~$\theta'$. On the contrary, the turning point $\zeta=\zeta_{1}$ of $\zeta^{\prime}$ is a turning point 
of $\theta'$ at $\theta=\pm\cos^{-1}\sqrt{\zeta}_{1}$. These considerations permit us to deduce that the shape of the string is folded.

Let us consider the branch of (\ref{zeta}) where $0\leqslant\theta\leqslant\pi/2$; the discussion on the branch where $0\leqslant\theta\leqslant\pi/2$ is utterly analogous. 
The solution~(\ref{solutionI1}) states that the string is partially stretched along the meridian of~$\mathrm{S}_{\lambda}^{2}$ at longitude~$\gamma=\omega\tau$ which corresponds 
to the interval $0\leqslant\sigma \leqslant\bar{\sigma}$. The string extends from \mbox{$\theta(0)=\cos^{-1}\sqrt{\zeta_{1}}\equiv\theta_{0}$} to the pole at~\mbox{$\theta(\bar{\sigma})=0$}. 
Since $\theta(\bar{\sigma})=0$ is not a turning point of~$\theta'$, in order to continue~$\theta$ in~\mbox{$\bar{\sigma}\leqslant\sigma\leqslant2\bar{\sigma}$}, we need the shift $\gamma\mapsto\gamma+\pi$.
Indeed, it follows from the embedding coordinates (\ref{embedding}) that~\mbox{$\theta\mapsto-\theta$} is equivalent to $\gamma\mapsto\gamma+\pi$. 
Thus, in the interval~\mbox{$\bar{\sigma}\leqslant\sigma \leqslant2\bar{\sigma}$}, the string extends from~\mbox{$\theta(\bar{\sigma})=0$} to~$\theta(2\bar{\sigma})=\theta_{0}$ at~$\gamma+\pi$, 
where $\theta'$ reaches the turning point $\theta_{0}$. The angle~$\theta$ along~\mbox{$2\bar{\sigma}\leqslant\sigma\leqslant4\bar{\sigma}$} follows the same path we have described backward. 
Due to the periodic boundary conditions~(\ref{bc2}), the angle~$\theta$ must cover the path over~$0\leqslant\sigma\leqslant4\bar{\sigma}$ an integer number of times, 
and~$\theta(\sigma)=\theta(2\pi-\sigma)$ holds as a consequence. The spinning string is folded as we have portrayed schematically in figure \ref{foldedpicture}, and the worldsheet is a strip rather than a~cylinder. 

\begin{figure}[H] 
\centering
\subfloat[]{\includegraphics[width=0.45\textwidth]{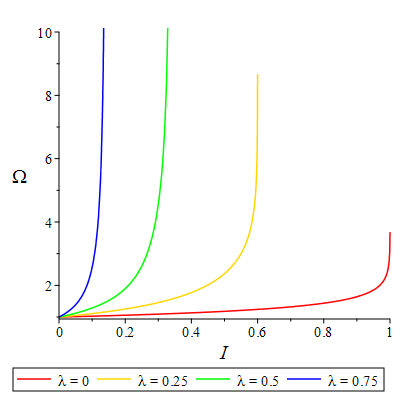}
\label{omegaI}}
\subfloat[]{\includegraphics[width=0.45\textwidth]{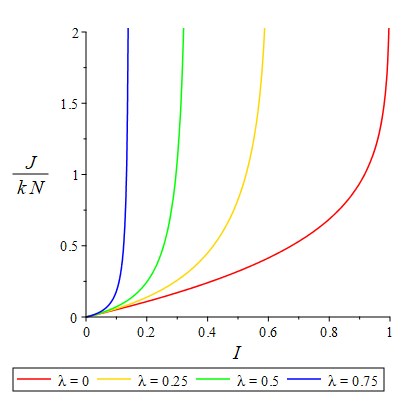}\label{JIplot}}
\caption{The quantities $\displaystyle \Omega=\omega/ N$ and $J / kN$ of folded strings through $0<I<I_{1}$ for some indicative values of $0\leqslant\lambda<1$. } 
\end{figure}

According to the previous discussion, (\ref{bc2}) implies that $\sigma$ must cover $0\leqslant\sigma\leqslant\bar{\sigma}$ a multiple of four number of times. Therefore, we deduce the quantization condition 
\begin{equation} 
\label{qc} 
4\bar{\sigma}N=2\pi \ ,
\end{equation}
where  the positive integer $N$ denotes the mode number that counts how many times the string is folded. It follows from (\ref{solutionI1}) and (\ref{qc}) that
\begin{equation}
\label{NI} 
\begin{split}
\Omega\equiv\frac{\omega}{N}&=\frac{2}{\pi}\left(\frac{1+\lambda}{1-\lambda}\right)^2\left[\,\K(m)-\frac{4\lambda}{1-4\lambda I-\lambda^2}\left(\frac{1-\lambda}{1+\lambda}-I\right)\Pi(n,m)\,\right] \ ,
\end{split}
\end{equation} 
where we have introduced $\Omega$ for later convenience and we have used that $\zeta=\zeta_{1}$ implies~\mbox{$\varphi=\pi/2$} in (\ref{argumentsI}) to reduce the incomplete elliptic integrals 
to complete elliptic integrals. We plot~$\Omega$ against the first integral $I$ in figure~\ref{omegaI}. The right-hand side of~(\ref{NI}) is analytic and real. Figure~\ref{omegaI} shows that it is also positive 
and strictly increasing. At the left endpoint, we have 
\begin{equation}
\label{NIleft}
\Omega\rightarrow1 \ , \quad I\rightarrow0 \ . 
\end{equation} 
We must exclude the value $I=0$ since it implies $\zeta=1$, that is, either $\theta=0$ or~$\theta=\pi$, where~$\gamma=\omega\tau$ is not defined. At the right endpoint, we have 
\begin{equation} 
\label{NIright}
\Omega\rightarrow\infty \ , \quad I\rightarrow I_{1}^{-} \ .
\end{equation}
The divergence follows from that of $\K(m)$ and $\Pi(n,m)$ at $m=1$ and $n=1$.

The angular momentum~$J$ implicitly determines the first integral $I$ by means of~(\ref{J0}). To perform the integration involved in~(\ref{J0}), we first split the interval 
$0\leqslant \sigma \leqslant 2\pi$ into the~$4N$ consecutive intervals $n\bar\sigma\leqslant\sigma\leqslant(n+1)\bar\sigma$, where $n=0,1,\dots, 4N-1$. We then use that each interval gives the same contribution, 
to equal the result of the integration to the integral over $0\leqslant\sigma\leqslant\bar\sigma$ multiplied by $4 N$. We finally use the ODE (\ref{ode}) to rephrase the integral. The result is 
\begin{equation}
\label{JI}
\begin{split}
\frac{J}{kN}&=\frac{1-\lambda^2}{2\pi\lambda}\int_{\zeta_{1}}^{1}\dif x\sqrt{\frac{1-x}{P(x)}}=\frac{2(1+\lambda)}{\pi(1-\lambda)}\left[\K(m)-\frac{(1+\lambda)^2}{1-4\lambda I-\lambda^2}\left(\frac{1-\lambda}{1+\lambda}-I\right)\Pi(n,m)\right] \ .
\end{split} 
\end{equation} 
We plot $J/kN$ against $I$ in figure \ref{JIplot}. In parallel with (\ref{NI}), the right-hand side of (\ref{JI}) is analytic and real, and figure \ref{JIplot} shows that it is also positive and strictly increasing. 
Hence, $J$ and $N$ have the same sign, and $J$ is positive. The ratio~$J /kN$ behaves like
\begin{equation}
\label{JIleft}
\frac{J}{kN}\rightarrow 0 \ , \quad I\rightarrow 0 \ ,
\end{equation} 
at the left endpoint, which is consistent with the fact $\gamma$ is not defined at the poles of $\mathrm{S}_{\lambda}^{2}$. At the right endpoint, we have
\begin{equation} 
\label{JIright}
\frac{J}{kN}\rightarrow \infty \ , \quad I\rightarrow I_{1}^{-} \ ,
\end{equation}
which follows from the divergence of $\K(m)$ and $\Pi(n,m)$ at $m=1$ and $n=1$. 

\subsubsection{Circular strings} 
\label{ssscircular} 

Let us analyze the class of spinning strings with $I>I_{1}$, where $0\leqslant\zeta\leqslant1$. Our starting point is the ODE (\ref{ode}) again. 
If we use the invariance under shifts of (\ref{ode}) to set $\zeta(0)=0$, we must choose the positive branch of (\ref{ode}). If $0\leqslant\sigma\leqslant\bar{\sigma}$ is the interval 
between two turning points of $\zeta'$, the ellipsoidal coordinate goes from $\zeta(0)=0$ to $\zeta(\bar{\sigma})=1$. To finally integrate (\ref{ode}), we must discriminate between $I_{1}<I<I_{2}$, 
in which case \mbox{$C>0$} and $\zeta_{1}<0$, and~$I>I_{2}$, where~\mbox{$C<0$} and $\zeta_{1}>\zeta_{2}$. If we assume $I_{1}<I<I_{2}$, we obtain 
\begin{equation} 
\label{solutionII1}
\begin{split}
& \omega\sigma=\int_{0}^{\zeta}\dif x\sqrt{\frac{\zeta_{2}- x}{P(x)}}  \\
&=\frac{\sqrt{1-\lambda^2}}{(1-4\lambda I -\lambda^2)\sqrt{I}}\left[ (1-\lambda)\F(\varphi,m)-\frac{4\lambda(1+\lambda)}{(1-\lambda)^2}
\left(I-\frac{1-\lambda}{1+\lambda}\right)\Pi(\varphi,n,m) 
\right] \ , 
\end{split}
\end{equation}
where
\begin{equation} 
\label{argumentsII}
\varphi=\sin^{-1}\sqrt{\frac{(1-\zeta_{1})\,\zeta}{\zeta-\zeta_{1}}} \ , \quad m=\sqrt{\frac{1-\lambda}{1+\lambda}\frac{1}{I}} \ , \quad n=\frac{1-4\lambda I-\lambda^2}{(1-\lambda)^2I} \ , 
\end{equation}
If we solved (\ref{ode}) for $I>I_{2}$, we would obtain (\ref{solutionII1}) again, whereby we deduce that (\ref{solutionII1}) is the classical solution if $I>I_{1}$.~\footnote{
Formula (\ref{solutionII1}) seems to have a simple pole at $I=I_{2}$, since $1-4\lambda I_{2}-\lambda^2=0$. However, this is not the case, but (\ref{solutionII1}) at $I=I_{2}$ reads
\begin{equation*}
\omega\sigma=\frac{2\sqrt{\lambda}(1+\lambda)}{\pi(1-\lambda)^2}\F(\,\sin^{-1}\sqrt\zeta,2\sqrt{\lambda}/(1+\lambda)\,) \ .
\end{equation*} 
To derive this expression we have used that $n=0$ and $\Pi(\varphi,0,m)=\F(\varphi,m)$ at $I=I_{2}$. This equation can be inverted to express $\zeta$ in terms of the Jacobian elliptic sine.} 
Note in passing that the mapping 
\begin{equation}
\sin\varphi\mapsto\frac{1}{\sin\varphi} \ , \quad m\mapsto\frac{1}{m} \ , \quad n\mapsto \frac{1}{n} \ . 
\end{equation} 
turns the arguments (\ref{argumentsII}) of circular strings into the arguments~(\ref{argumentsI}) of folded strings. 

We could construct the solution to (\ref{ode}) for any interval other than $0\leqslant\sigma\leqslant\bar{\sigma}$ by alternating the branches of (\ref{ode}). 
The procedure parallels the discussion in subsection~\ref{sssfolded}, and we will not repeat it. Instead, we look at the shape of the string described by (\ref{solutionII1}). 
Since $\zeta=0$ and~$\zeta=1$ are turning points of $\zeta'$ but not of $\theta'$, we can deduce that the string circular. The line of thought is analogous to its counterpart in subsection \ref{sssfolded}. 

Let us consider the branch of (\ref{zeta}) where $0\leqslant\theta\leqslant\pi/2$ . Solution~(\ref{solutionI1}) states that the string is totally stretched along 
the meridian of~$\mathrm{S}_{\lambda}^{2}$ at longitude $\gamma=\omega\tau$ that corresponds to~\mbox{$0\leqslant\sigma \leqslant\bar{\sigma}$}. 
The string extends from the point of the equator at $\theta(0)=\pi/2$ to the pole at~\mbox{$\theta(\bar{\sigma})=0$} along the interval~$0\leqslant\sigma\leqslant\bar{\sigma}$. 
The point $\theta(\bar{\sigma})=0$ is not a turning point of~$\theta'$. The shift~\mbox{$\gamma\mapsto\gamma+\pi$} allows us to continue $\theta'$ 
along~$\bar{\sigma}\leqslant\sigma\leqslant2\bar{\sigma}$. The string extends from~$\theta(\bar{\sigma})=0$ to~$\theta(2\bar{\sigma})=\pi/2$ at longitude~$\gamma+\pi$. 
The point~$\theta(2\bar{\sigma})=\pi/2$ is not a turning point of~$\theta'$ either. To continue~$\theta$, we need to consider the branch of (\ref{zeta}) where~$\pi/2\leqslant\theta\leqslant\pi$. 
The angle~$\theta$ then follows a similar path along~$2\bar{\sigma}\leqslant\sigma\leqslant4\bar{\sigma}$ to the one we have described above, 
where~$\theta(3\bar{\sigma})=\pi$ and~\mbox{$\theta(4\bar{\sigma})=\pi/2$}. Therefore,~$\theta'$ never reaches a turning point, and the string surrounds 
a great circle of~$\mathrm{S}_{\lambda}^{2}$ that is perpendicular to the equator. Periodic boundary conditions~(\ref{bc2}) imply that~$\theta$ must cover 
the great circle over~$0\leqslant\sigma\leqslant4\bar{\sigma}$ an integer number of times. It follows from this discussion that the shape of the string is circular 
as we have depicted schematically in figure \ref{circularpicture} above. 

\begin{figure}[H] 
\centering
\subfloat[]{\includegraphics[width=0.45\textwidth]{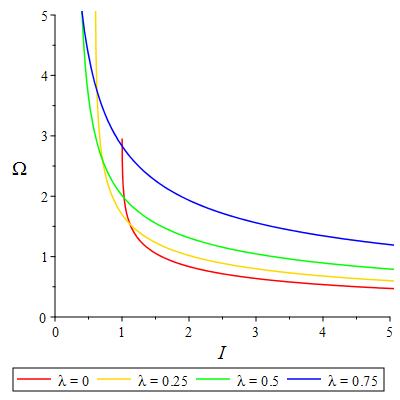}\label{omegaII}}
\subfloat[]{\includegraphics[width=0.45\textwidth]{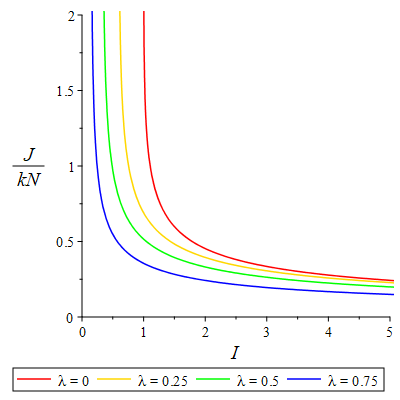}\label{JIIplot}}
\caption{
The quantities $\displaystyle \Omega=\omega/N$ and $J/kN$ of folded strings through  $I_{1}<I<\infty$ for some indicative values of $0\leqslant\lambda<1$. 
For different values of $\lambda$, the curves of $\Omega$ intersect.}
\end{figure}

It follows from our previous discussion that periodicity implies a quantization condition, which is formally identical to the equation (\ref{qc}). Nonetheless, the mode number~$N$ 
is now the positive integer that counts how many times the string wraps a great circle of~$\mathrm{S}_{\lambda}^{2}$. The quantization condition implies that
\begin{equation}
\label{NII} 
\Omega\equiv\frac{\omega}{N}=\frac{2\sqrt{1-\lambda^2}}{\pi(1-4\lambda I -\lambda^2)\sqrt{I}}\left[(1-\lambda)\K(m)-\frac{4\lambda(1+\lambda)}{(1-\lambda)^2}
\left(I-\frac{1-\lambda}{1+\lambda}\right)\Pi(n,m)
\right] \ . 
\end{equation}
We plot $\Omega$ against $I$ in figure~\ref{omegaII}. The right-hand side of (\ref{NII}) is analytic and real. Figure~\ref{omegaII} shows that it is positive and strictly decreasing. 
This behavior contrasts with~$\Omega$ for folded strings, which is positive but strictly increasing with respect to $I$. We have
\begin{equation}
\label{NIIleft} 
\Omega\rightarrow\infty \ , \quad I\rightarrow I_{1}^{+} \ . 
\end{equation} 
and
\begin{equation} 
\label{NIIright}
\Omega\rightarrow0 \ ,\quad I\rightarrow \infty \ , 
\end{equation} 
at the left and right endpoints, respectively. 

The angular momentum $J$ implicitly fixes the first integral $I$ through (\ref{J0}) like in the case of folded strings. To perform the integral in the right-hand side of (\ref{J0}), 
we must discriminate between the ranges~$I_{1}< I<I_{2}$ and $I>I_{2}$ again. Let us assume that~$I_{1}<I<I_{2}$. 
If we rephrase the integral over~$0\leqslant\sigma\leqslant2\pi$ into $4N$ times the integral over~$0\leqslant\sigma\leqslant\bar{\sigma}$, in parallel with the previous subsection, we obtain 
\begin{equation}
\label{JII1}
\begin{split}
\frac{J}{kN}&
=\frac{1-\lambda^2}{2\pi\lambda}\int_{0}^{1}\dif x\sqrt{\frac{1-x}{P(x)}} \\
&=\frac{2\sqrt{1-\lambda^2}}{\pi(1-4\lambda I-\lambda^2)}\left[(1-\lambda)\sqrt{I}\K(m)-\frac{(1+\lambda)^2}{(1-\lambda)\sqrt{I}}\left(I-\frac{1-\lambda}{1+\lambda}\right)\Pi(n,m)\right] \ .
\end{split}
\end{equation}
Since the integral of (\ref{J0}) with $I>I_{2}$ is (\ref{JII1}) again,~(\ref{JII1}) is the correct expression of the angular momentum $J$ for every $I>I_{1}$.
\footnote{
Similarly to (\ref{solutionII1}), formula (\ref{JII1}) seems to have a simple pole at $I=I_{2}$. 
However, $I=I_{2}$ is just a removable singularity of (\ref{JII1}), where the angular momentum reads
\begin{equation*}
\frac{J}{kN}=\frac{4\sqrt{\lambda}}{(1-\lambda)\pi}\K(\,2\sqrt{\lambda}/(1+\lambda)\,) \ .
\end{equation*} 
}
We plot~\mbox{$J/kN$} against $I$ in figure \ref{JIIplot}. The ratio $J/N$ is analytic and real. Figure \ref{JIIplot} shows that it is also positive and strictly decreasing. 
The pair $J$ and $N$ have the same sign. At the endpoints, we have
\begin{equation}
\label{JIIleft}
\frac{J}{kN}\rightarrow\infty \ , \quad I\rightarrow I_{1}^{+} \ , 
\end{equation}
and 
\begin{equation}
\label{JIIright}
\frac{J}{kN}\rightarrow0 \ ,\quad I\rightarrow \infty \ .
\end{equation} 

\subsubsection{The dispersion relation} 
\label{sssdisp} 

We will consider now the dispersion relation of both folded and circular strings. The dispersion relation is the expression of the energy $E$ in terms of the angular momentum~$J$ as well as the mode number~$N$. 
To highlight the connection between the dispersion relation and the semiclassical limit, we briefly sketch the r\^ole of spinning strings as solitons in the conformal field theory of the superstring worldsheet. 

We begin the discussion by recalling the connection between spinning strings and closed-superstring vertex operators. First of all, vertex operators must carry the quantum numbers $E$, $J$, and $N$ for the quantities to be meaningful at the quantum level. 
The symmetries associated to~$E$ and~$J$ must be non-anomalous and~$N$ must label different superselection sectors of the Hilbert space. 
Two-point functions of vertex operators admit a path-integral representation that realizes the transition amplitude between the ``initial'' vertex operator at~\mbox{$\tau=-\infty$} and the ``final'' vertex operator at~\mbox{$\tau=\infty$}. 
Spinning strings arise in the saddle-point approximation of this path integral in the guise of solitons \cite{0304139,1002.1716}. The \mbox{saddle-point} approximation that gives rise to spinning strings is applicable in the semiclassical limit of large level~$k\rightarrow\infty$. 
The conserved charges of spinning strings vanish unless they are semiclassical, that is, unless~$E,J\sim k\rightarrow\infty$ with $E/k$ and $J/k$ being fixed and finite. 
On the contrary, the semiclassical limit does not constrain~$N$. Since the marginality condition of vertex operators maps to the dispersion relation of spinning strings, the latter is the condition of classical conformal invariance in the superstring~worldsheet.

Let us consider the dispersion relation. To perform the computation, we must express~$\omega$ and~$I$, which parametrize classical trajectories, in terms of the quantities $J$ and~$N$ of spinning strings.  
We can complete this task by following the subsequent steps.

First, we will consider folded strings. The quantities~$J$ and $N$ are related to~$\omega$ and~$I$ via~(\ref{NI}) and~(\ref{JI}). Equation~(\ref{JI}) fixes  $I=I(J/kN)$ implicitly. 
Figure~\ref{JIplot} shows that the correspondence between $0<I<I_{1}$ and~$0<J/kN<\infty$ is one-to-one and onto, and~\mbox{$I=I(J/kN)$} is single-valued for $0\leqslant\lambda<1$ 
as a consequence. Given $I(J/kN)$, equation~(\ref{NI}) determines $\omega=N\Omega(J/kN)$. Figure \ref{omegaI} shows that the correspondence between~\mbox{$0<I<I_{1}$} and $1<\Omega<\infty$ 
is one-to-one and onto. Therefore, $\Omega(J/kN)$ is \mbox{single-valued}. The Virasoro constraint (\ref{disp}) and formula~(\ref{E0}) permit to write $E$ in terms of~$I$ and~$\Omega$, 
and, hence, $E$ in terms of $N$ and~\mbox{$J/N$}. 

These steps also apply to circular strings. Equations~(\ref{NII}) and~(\ref{JII1}) determine~$I(J/kN)$ and $\omega=N\Omega(J/kN)$. Figures \ref{omegaII} and \ref{JIIplot} 
show that $I(J/kN)$ and~$\Omega(J/kN)$ are \mbox{single-valued}. Formulae (\ref{disp}) and~(\ref{E0}) permit us to write $E$ in terms of~$N$ and $J/kN$.

\begin{figure}[H]
\centering
\subfloat[]{\includegraphics[width=0.45\textwidth]{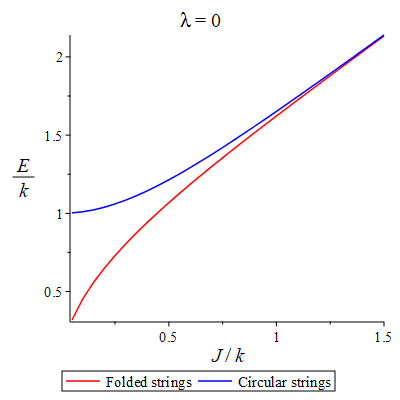}\label{disp1}}
\subfloat[]{\includegraphics[width=0.45\textwidth]{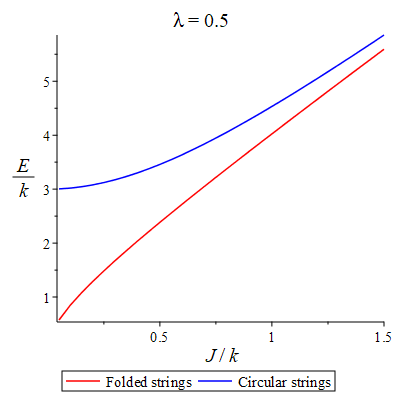}\label{disp2plot}}

\caption{The dispersion relation of spinning strings with $N=1$ for $0\leqslant\lambda<1$. Folded strings are energetically favourable. The $\lambda$-deformation increases the overall energy and widens the gap between folded and circular strings.}
\end{figure}

The formula for the dispersion relation has the form
\begin{equation}
\label{disp2}
E=k N\Omega(J/kN)\sqrt{\frac{1+\lambda}{1-\lambda}I(J/kN)} \ ,
\end{equation}
where we have chosen $E$ to be positive in relating the energy and the Virasoro constraints.
In general, (\ref{disp2}) cannot be computed analytically. We will address the analytic computation for nearly degenerate spinning strings in subsection \ref{ssndeg}. 
In this subsection, we will evaluate numerically (\ref{disp2}). We plot the results with~$N=1$ in figures~\ref{disp1} and~\ref{disp2plot}. There is no loss of generality 
in the choice~\mbox{$N=1$} because~(\ref{disp2}) just involves~\mbox{$E/N$} and $J/N$.

Figures \ref{disp1} and \ref{disp2plot} show that folded strings are energetically favorable for given $J$ throughout~\mbox{$0\leqslant \lambda<1$}. 
They also make clear that the effect of the $\lambda$-deformation is twofold. First, the $\lambda$-deformation increases $E$ for both folded and circular strings. Second, the \mbox{$\lambda$-deformation} 
enlarges the gap between the energies of circular and folded strings. We will see in later than in the NATD limit, where $\lambda\rightarrow1$, just folded strings remain since the energetic gap becomes infinitely wide. 

One may make plausible the predominance of folded strings by the following argument. The energy and the angular momentum are proportional to the tension for both folded and circular strings. 
Therefore, the greater the tension of spinning strings is, the faster they rotate. While an increment of tension pulls the endpoint folded strings, circular strings are always totally stretched. 
Given the angular speed, the contribution of the stretching to the energy of spinning strings is always greater for circular strings than for folded strings. 
For this reason, folded strings should always have less energy than circular strings. 

\subsection{Point particles and nearly degenerate spinning strings} 
\label{ssndeg}

In this subsection, we will focus on spinning strings whose dispersion relation is analytically computable. 
In particular, we will first consider point particles, and turn to nearly degenerate spinning strings afterward. 
We will find the corresponding the dispersion relation analytically 
at leading order in $J$.~\footnote{
In subsections \ref{sssnpoint}--\ref{sssslow}, we will base the discussion on limits of the angular momentum $J$ for brevity. 
However, one must keep in mind that the quantity actually involved in the limits is $J/k$ rather than $J$.
}

\subsubsection{Point particles} 
\label{ssspoint}

Point particles are degenerate spinning strings with $N=0$, where periodicity trivially holds. 
Since $\theta^{\prime}=0$, we must check again that the ansatz~(\ref{ansatz2}) consistently truncates the equations of motion of the action (\ref{S}). The only consistent choices are, on the one hand,~$\theta=0$ and $\theta=\pi$, and, on the other, $\theta=\pi/2$.

If either~$\theta=0$ and $\theta=\pi$, the angle $\gamma$ is not defined. We deduce from the Virasoro constraint~(\ref{V1}), the energy (\ref{E0}), and the angular momentum  (\ref{J0}) that $E=J=0$. Being localized at a pole of~$\mathrm{S}_{\lambda}^{3}$ and $t=0$, 
we can deem point particles with vanishing energy to be instantons. Actually, we can place the instanton anywhere in $\mathrm{S}^2_\lambda$, not only at the poles. 

Moreover, the value~$\theta=\pi/2$ corresponds to BMN particles, that is, point particles that surround the great circle of
of~$\mathrm{S}_{\lambda}^{2}$. Equations~(\ref{V1}), (\ref{E0}), and (\ref{J0}) imply that the dispersion relation of BMN particles is
\begin{equation}
\label{EBMN}
E=\frac{1+\lambda}{1-\lambda}J \ . 
\end{equation}
We conclude that effect of the $\lambda$-deformation for BMN particles is the overall increment of the energy by \mbox{$(1+\lambda)/(1-\lambda)$}.

\subsubsection{Nearly point-like strings}
\label{sssnpoint}

Nearly point-like strings are folded strings with either small $J$ or large $N$. Equation~(\ref{JIleft}) implies $J/N\rightarrow 0$ when $I\rightarrow0$. Since $I\rightarrow0$ 
implies $\zeta_{1}\rightarrow1$, the interval $\zeta_{1}\leqslant\zeta\leqslant1$ tends to  collapse into a point, and, thus, strings are ``nearly point-like". Nearly point-like strings with $\lambda=0$ and $N=1$ are considered in~\cite{1311.5800}. 

First, we consider the classical solution. If we choose the branch of  (\ref{zeta}) that corresponds to~$0\leqslant\theta\leqslant\pi/2$, we obtain
\begin{equation} 
\label{thetanpoint}
\theta=\sqrt{2\frac{1-\lambda}{1+\lambda}I}\, |\cos(\omega\sigma)|+\mathrm{O}(I) \ .
\end{equation}
To write (\ref{thetanpoint}), we have performed a series expansion around $I=0$ everywhere in (\ref{solutionI1}) with~$\varphi$ fixed, and, in addition, we have employed
\begin{equation}
\label{Fm0}
\F(\varphi,0)=\varphi \ , \qquad \Pi(\varphi,0,0)=\varphi \ .
\end{equation}
The solution (\ref{thetanpoint}) implies that the angle $\theta$ oscillates around the pole of $\mathrm{S}^{2}_{\lambda}$ at $\theta=0$ 
with small amplitude~$\theta_{0}=\sqrt{((1-\lambda)I)/(1+\lambda)}$. In the extreme $\theta_{0}=0$, the string collapses at $\theta=0$. Moreover,~(\ref{thetanpoint}) is consistent with (\ref{NIleft}), 
which states that~\mbox{$\omega\rightarrow N$} as~$I\rightarrow0$. 

The angle $\theta$ in (\ref{thetanpoint}) mimics the radial coordinate $r$ of folded strings in flat space (see, for instance, equation~(3.2) of \cite{1012.3986}). 
The dispersion relation supports the analogy. Performing the series expansion of (\ref{NI}) and (\ref{JI}) around~$I=0$, we obtain that
\begin{align}
\Omega=1+\mathrm{O}(I) \ , \\
\frac{J}{k N}=\frac{1}{2}I+\mathrm{O}(I^2) \ ,
\end{align}
which we can use to show that (\ref{disp2}) indeed obeys the Regge dispersion relation
\begin{equation}
\label{Enpoint}
E=\sqrt{\frac{1+\lambda}{1-\lambda}2kNJ} +\mathrm{O}(J) \ .
\end{equation}
One may justify the appearance of Regge trajectories by arguing that effects of the curvature of target space are suppressed for nearly point-like strings, and, hence, the behavior of the $\lambda$-deformation of nearly point-like strings should be roughly the same as strings in flat space. 
We see that effect of the $\lambda$-deformation is indeed limited to the overall increment of the energy by~\mbox{$\sqrt{(1+\lambda)/(1-\lambda)}$}.

\subsubsection{Fast spinning strings}  
\label{sssfast}

Fast string are folded and circular strings with large $J$. Equation~(\ref{JIright}) implies that~\mbox{$J\rightarrow\infty$} when~$I\rightarrow I_{1}^{-}$ 
for folded strings and (\ref{JIIleft}) implies $J\rightarrow\infty$ when $I\rightarrow I_{1}^{+}$ for circular strings. Note that $J/N\rightarrow\infty$ 
cannot be reached by tuning $N$. Formulae~(\ref{NIright}) and~(\ref{NIIleft}) state that~$\omega\rightarrow\infty$ when $I\rightarrow I_{1}$, and, hence, 
the rotation of spinning strings in this limit is ``fast". Fast strings with~$\lambda=0$ and~$N=1$ are considered in~\cite{1311.5800}. 

We will begin with the dispersion relation of folded string. We must use (\ref{JI}) to compute the series of $I=I(J/kN)$ around~$J=\infty$ and then 
use the result in~(\ref{NI}) to obtain the series of $\Omega=\Omega(J/kN)$ around $J=\infty$. We deduce from (\ref{argumentsI}) that $m=n=1$ hold when~\mbox{$I=I_{1}$}. 
Therefore, we need the asymptotic series of~$\K(m)$ around $m=1$
\begin{equation}
\label{serie1}
\K(m)\simeq-\frac{1}{2}\ln(1-m^2)\ , \quad m\rightarrow1 \ .
\end{equation}
We also need the asymptotic series of  $\Pi(n,m)$ around $m=n=1$. If either $m=1$ or~$n=1$ holds, $\Pi(n,m)$ diverges. However, the asymptotic series of~$\Pi(n,m)$ around~$I=I_{1}$ 
depends on the specific dependence of $m$ and $n$ on $I$ in a quite involved series. To compute the dispersion relation at leading order when $J\rightarrow\infty$ in this case, 
we just need to know that~$(I_{1}-I)\Pi(n,m)$ is subleading because it is non-vanishing and finite when~$I\rightarrow I_{1}^{+}$. This information can be inferred 
from numerical evaluation of $(I_{1}-I)\Pi(n,m)$. If we ignore~$(I_{1}-I)\Pi(n,m)$ in (\ref{JI}), it follows from (\ref{serie1}) that
\begin{equation}
\label{Ifast}
I\simeq  \frac{1-\lambda}{1+\lambda}\left(1-\exp(-\frac{(1-\lambda)\pi J}{(1+\lambda)kN})
\right) \ , \quad J\rightarrow\infty \ .
\end{equation}
Therefore, (\ref{NI}) implies that
\begin{equation}
\label{omegafast}
\Omega\simeq\frac{(1+\lambda)J}{(1-\lambda)kN} \ , \quad J\rightarrow\infty \ .
\end{equation}
Both (\ref{Ifast}) and (\ref{omegafast}) are also valid for circular strings. They follow from the leading order of the series of (\ref{NII}) and~(\ref{JII1}) around $J\rightarrow\infty$.

If we introduce both (\ref{Ifast}) and (\ref{omegafast}) into (\ref{disp2}), we are led to
\begin{equation}
\label{Efast}
E\simeq\frac{1+\lambda}{1-\lambda}J \ , \quad J\rightarrow\infty \ .
\end{equation}
The dispersion relation is linear, consistently with the behavior of figures \ref{disp1} and \ref{disp2plot} at large~$E$ and~$J$. (To obtain the intercept of the line of (\ref{Efast}) in figures \ref{disp1} 
and \ref{disp2plot}, we would need the series of $(I_{1}-I)\Pi(n,m)$ around $J=\infty$.)  The effect of the $\lambda$-deformation is the overall increment of the energy 
by $(1+\lambda)/(1-\lambda)$, similarly to (\ref{EBMN}). It is worth stressing that no logarithmic divergences occur, just as in the case $\lambda=0$~\cite{1311.5800}.

Let us now turn to the classical solution.  We will choose the branch of~(\ref{zeta}) where \mbox{$0\leqslant\theta\leqslant\pi/2$} for definiteness and start from folded strings. 
Since $I\rightarrow I_{1}^{-}$ implies~$\zeta_{1}\rightarrow 0$, the interval $\zeta_{1}\leqslant\zeta\leqslant1$ approaches $0\leqslant\zeta\leqslant1$. 
Nonetheless, if we had applied $I\rightarrow I_{1}^{-}$ directly to~(\ref{solutionI1}), we would have obtained a divergent result. Furthermore, the series expansion 
around~$I=I_{1}$ of (\ref{solutionI1}) is quite involved. To clarify the behavior of fast strings, we will apply~$I=I_{1}^{-}$ instead to the right-hand side of (\ref{solutionI1}) with~$0\leqslant\varphi< \pi/2$ fixed. Since
\begin{equation}
\F(\varphi,1)=\tanh^{-1}(\sin\varphi) \ ,
\end{equation}
and 
\begin{equation}   
\underset{I\rightarrow I_{1}^{-}}{\lim}\left(\frac{1-\lambda}{1+\lambda}-I\right)\Pi(\varphi,n,m)=0 \ ,
\quad \mathrm{with} \quad 0\leqslant\varphi<\frac{\pi}{2} \quad \mathrm{fixed} \ ,
\end{equation}
we obtain
\begin{equation}
\label{thetafast}
\cos^2\theta=\frac{\zeta_{1}}{1+\zeta_{1}\tanh^2(\bar{\omega}\sigma)} \ ,
\end{equation}
where $\bar{\omega}\equiv[(1-\lambda)^2\omega]/(1+\lambda)^2$. Note that the classical solution~(\ref{thetafast}) also realizes the fast-string limit of circular strings approximately.

Based on (\ref{thetafast}), one can now present the following picture. The limit $I\rightarrow I_{1}^{-}$ pulls the endpoints of the folded string at latitude $\theta=\cos^{-1}\sqrt{\zeta_{1}}$ 
toward the equator at $\theta=\pi/2$.  It also increases the angular speed because $\omega\rightarrow\infty$. The string stretches and gains speed until it reaches the equator, 
where periodicity cannot hold anymore, rotations are infinitely fast, and the string disintegrates. Moreover, circular strings increase their angular speed as $I\rightarrow I_{1}^{+}$, 
and they disintegrate when rotations are infinitely fast.

Let us finally note that we can deem the BMN particles of subsection \ref{ssspoint} to be the threshold between folded and circular strings, in the sense that they constitute 
the only class of solutions realized by the ansatz (\ref{ansatz2}) that admit $J=\infty$ for $\theta=\pi/2$.

\subsubsection{Slow spinning strings}

\label{sssslow}

Slow strings are circular strings with either small $J$ or large~$N$.  It follows from (\ref{JIIright}) that the limit~$J/N\rightarrow 0$ corresponds to~$I\rightarrow\infty$. 
Since (\ref{NIIright}) implies that $\omega/N\rightarrow0$ when~\mbox{$I\rightarrow\infty$}, the rotations of circular strings are ``slow" in this limit. In a sense, slow strings are the counterpart of nearly point-like strings of subsection \ref{sssnpoint} in the class of circular strings. 
Slow strings with~$\lambda=0$ and $N=1$ have been considered before in \cite{1311.5800}.

Let us begin with the classical solution. If we perform the series expansion of the solution (\ref{solutionII1}) around $I=\infty$, we obtain
\begin{equation}
\label{thetaslow}
\mathrm{cotan}^2\theta=\tan^2\left(\sqrt{\frac{1+\lambda}{1-\lambda}I}\,\omega\sigma\right) +\mathrm{O}(I^{-1})\ .
\end{equation}
To obtain (\ref{thetaslow}), we have employed (\ref{Fm0}) and
\begin{equation}
\Pi(\varphi,n,0)=\frac{\tan^{-1}(\sqrt{1-n}\tan\varphi)}{\sqrt{1-n}} \ .
\end{equation}
From equation (\ref{thetaslow}) we conclude that $\theta$ ranges along a great circle of $\mathrm{S}^{2}_{\lambda}$ with trigonometric periods, which are nothing but degenerate elliptic periods. 
Trigonometric periods appear just before the string ceases to spin at~$I=\infty$, where it disintegrates.

It follows from either (\ref{thetaslow}) and the series of (\ref{NII}) around $I=\infty$ that 
\begin{equation}
\label{omegaslow}
\Omega=\sqrt{\frac{1+\lambda}{(1-\lambda)I}}+\mathrm{O}(I^{-3/2}) \ .
\end{equation}
In addition, the series expansion for the angular momentum~(\ref{JII1}) around $I=\infty$ reads
\begin{equation}
\label{Jslow}
\frac{J}{ kN} = \frac{1}{2}\sqrt{\frac{1-\lambda^2}{I}}+\mathrm{O}(I^{-3/2}) \ .
\end{equation}
If we introduce (\ref{omegaslow}) and (\ref{Jslow}) in the formula of the dispersion relation (\ref{disp2}), we obtain
\begin{equation}
\label{Eslow} 
E=\frac{1+\lambda}{1-\lambda}k\abs{N} + \mathrm{O}\left(\frac{J^{2}}{N^2}\right) \ . 
\end{equation}
Formula (\ref{Eslow}) displays harmonic dependence, consistently with the expectations. The effect of the $\lambda$-deformation is the overall increment 
of the energy by $(1+\lambda)/(1-\lambda)$, similarly to (\ref{EBMN}) and (\ref{Efast}). However, this factor does not appear in front of the angular momentum~$J$. Note also that the leading order of the dispersion relation becomes independent of $J$. 


\section{The non-Abelian T-dual limit and the spinning-string ansatz}
\label{sintNATD}

In this section, we will present the NATD limit, where $\lambda=1$. In subsection~\ref{ssactionNATD}, we will review the NATD limit of the \mbox{$\lambda$-deformation} of the $\SU$ WZW model. 
We will present the coordinate systems, the target-space fields, and the classical action of the model. In subsection~\ref{ssansatzNATD}, we will introduce the spinning-string ansatz in the NATD limit. 
Throughout this section, we will focus on the NATD limit directly and comment on any connection with the regime $0\leqslant\lambda<1$ when necessary.

\subsection{The classical action}

\label{ssactionNATD}

First, we must recall relation (\ref{lambda}), 
\begin{equation}
\label{lambdaNATD}
\lambda=\frac{k}{1+k} \ ,
\end{equation}
where, following \cite{1410.1886}, we have set $\kappa=1$ to simplify later formulae. The point $\lambda=1$ corresponds to the limit $k \rightarrow \infty$. 
The NATD limit not only consists of the application of $k\rightarrow\infty$, but also of the rescaling of the target-space coordinates~\cite{1312.4560}. 
The rescaling is necessary to obtain the series of $g\in\SU$ with respect to $1/k$ around~$g=1$ in the classical action. In particular, the NATD limit of the $\lambda$-deformation 
of the $\SU$ WZW model involves the rescaling~\cite{1312.4560}
\begin{equation}
\label{rescS}
\theta=\frac{r}{2k} \ , \quad \mathrm{with} \ \beta \  \mathrm{and} ~\gamma \ \mathrm{fixed} \ .
\end{equation}
The limit $k\rightarrow\infty$ zooms in the pole of $\mathrm{S}_{\lambda}^{3}$ at $\theta=0$, where the $\mathrm{S}^{2}$ parametrized by $\beta$ and~$\gamma$ shrinks to a point.  The NATD limit of the metric (\ref{GS}) and the NATD limit of the B-field (\ref{BS}) are
\begin{align}
\label{GSNATD}
\dif s^2&=\frac{1}{2}\left[\dif r^2+\frac{r^2}{r^2+1}(\dif \beta^2+\sin^2\beta\dif\gamma^2)\right] \ , \\ 
\label{BSNATD}
B&=-\frac{r^3}{2(r^2+1)}\sin\beta\dif\beta\wedge\dif\gamma \ ,
\end{align}
respectively. The limit $k\rightarrow\infty$ implies $0\leqslant r<\infty$, and $r$ parametrizes a non-compact radial direction. We will denote the NATD limit of $\mathrm{S}_{\lambda}^{3}$ by $\widetilde{\mathrm{R}}^{3}$ because it is topologically the three-dimensional Euclidean space.

As we have already mentioned in subsection \ref{ssaction}, we will assume that the type IIB supergravity background that embeds the $\lambda$-deformation of the $\SU$ WZW model is consistent with the NATD limit. 
This assumption implies that the NATD limit must be applicable to the supergravity fields. In particular, we will need the NATD limit of $t$. To specify the limit, we will follow \cite{1410.1886}, like in subsection \ref{ssaction}, and rescale
\begin{equation}
\label{tNATD}
t=\frac{\tilde{t}}{2 k} \ , 
\end{equation}
before we apply the NATD limit. We must note that the NATD limit does not modify the compactness of the time-like direction, thus $-\infty<\tilde{t}<\infty$.

Like in subsection \ref{ssaction}, the NATD limit of the RR three-form flux will have no components along~$\widetilde{\mathrm{R}}^{3}$. 
This statement directly follows from the assumption that it does not have components along $\mathrm{S}_{\lambda}^{3}$ prior to the NATD limit.
We will assume the use of S-duality again to interchange the B-field (\ref{BSNATD}) and the RR two-form potential of the type IIB superstring background in the NATD limit. In this way, we will be able to ignore the S-dual NSNS flux, and the spinning-string ansatz (\ref{ansatzNATD}) will consistently truncate the equations of motion.

We can now write action for closed bosonic strings through the truncation of the NATD limit of the S-dual background to $\widetilde{\mathrm{R}}^{3}$ with the time-like direction. In parallel with (\ref{S}), the classical action is the conformally gauge-fixed Polyakov action
\begin{equation}
\label{SNATD}
\begin{split}
S_{\mathrm{P}}&=-\frac{1}{8\pi}\int_{-\infty}^{\infty}\dif\tau\int_{0}^{2\pi}\dif\sigma\left(\dot{\tilde{t}}^2-\tilde{t}'^2-\dot{r}^2+r'^2\vphantom{\frac{r^2}{r^2+1}}\right.
\\&\left.+\frac{r^2}{r^2+1}[\,-\dot{\beta}^2+\beta'^2 + \sin^2\beta \,(-\dot{\gamma}^2+\gamma'^2)\,]\right)\ ,
\end{split}
\end{equation}
Again, the topological term is absent due to the condition we have imposed to the initial RR three-form flux.

We must supplement the classical action with the Virasoro constraints, namely
\begin{equation}
\label{V1NATD}
\dot{\tilde{t}}^2+\tilde{t}'^2=\dot{r}^2+r'^2+\frac{r^2}{r^2+1}[\,\dot{\beta}^2+\beta'^2 + \sin^2\beta \,(\dot{\gamma}^2+\gamma'^2) ] 
\end{equation}
and
\begin{equation}
\label{V2NATD}
\dot{\tilde{t}}\tilde{t}'=\dot{r}r'+\frac{r^2}{r^2+1}(\,\dot{\beta}\beta' + \sin^2\beta \,\dot{\gamma}\gamma') \ .
\end{equation}
Formulae (\ref{V1NATD}) and (\ref{V2NATD}) are the NATD limit of (\ref{V1}) and (\ref{V2}), respectively. We must also impose periodic boundary conditions to the target-space coordinates. 
The periodicity of~$t$ and $\theta$ implies the periodicity of $\tilde{t}$ and $r$. On the other hand, the periodic boundary conditions of~$\beta$ and $\gamma$ are written in (\ref{bc}). 
Finally, the invariance of the classical action under shifts of~$\tilde{t}$ and $\gamma$ again implies the conservation of the energy $E$ and the angular momentum $J$, respectively. 
Noether's theorem straightforwardly yields the expressions of these conserved charges. 

\subsection{The spinning-string ansatz}
\label{ssansatzNATD} 

If we consider the ansatz (\ref{ansatz2}) together with the rescaling (\ref{rescS}), we deduce that the spinning-string ansatz in $\widetilde{\mathrm{R}}^{3}$ is
\begin{equation}
\label{ansatzNATD}
\tilde{t}=\tilde{\mu}\tau \ , \quad r=r(\sigma) \ , \quad \beta=\frac{\pi}{2}  \ , \quad \gamma=\omega\tau \ .
\end{equation}
The parameter $\tilde{\mu}$ is the NATD limit of $\mu$, and they are related through $\mu=\tilde{\mu}/2k$. We will assume $\omega\geqslant0$ without loss of generality. 
The periodicity of~$\tilde{t}$,~$\beta$, and $\gamma$ trivially holds, whereas we must still supplement $r$ with
\begin{equation}
\label{bcNATD}
r(\sigma+2\pi)=r(\sigma) \ .
\end{equation} 

The ansatz (\ref{ansatzNATD}) describes spinning motion, but, unlike (\ref{ansatz2}), rotations take place in a non-compact space. 
More precisely, the ansatz specifies $\widetilde{\mathrm{R}}^{2}$ inside $\widetilde{\mathrm{R}}^{3}$, the NATD limit of~$\mathrm{S}_{\lambda}^{2}$. 
The space $\widetilde{\mathrm{R}}^{2}$ is topologically the real plane, where $r$ is the radius and $\gamma$ is the polar angle.
The ansatz~$\gamma=\omega\tau$ implies that strings are stretched along the radial rays and rotate around the center of~$\widetilde{\mathrm{R}}^{2}$ at~$r=0$. 
The ansatz~$r=r(\sigma)$ quantifies the extension of strings along radial rays. Note also that rotations are isochronous with respect to $\tilde{t}=\tilde{\mu}\tau$.

If we introduce (\ref{ansatzNATD}) in the equations of motion of (\ref{SNATD}), we obtain an ODE. This differential equation is the equation of motion of the Lagrangian of a mechanical system, which reads
\begin{equation}
\label{LNATD}
L = \frac{1}{4}\left(r'^2-\omega^2\frac{r^2}{r^2+1}\right) \ .
\end{equation}
Since the Lagrangian is a rational function of $r$, there is no need to introduce a coordinate like $\zeta$ in (\ref{zeta}). The equations of motion of (\ref{LNATD}) determine classical trajectories, 
and we must impose the Virasoro constraints to force them to realize spinning strings. The ansatz~(\ref{ansatzNATD}) reduces the first Virasoro constraint (\ref{V1NATD}) to
\begin{equation}
\label{dispNATD}
\frac{1}{4}\tilde{\mu}^2=H \ ,
\end{equation}
where $H$ is the mechanical energy of the system
\begin{equation}
\label{HNATD} 
H = \frac{1}{4}\left(r'^2+\omega^2\frac{r^2}{r^2+1}\right) \ .
\end{equation}
The ansatz~(\ref{ansatzNATD}) also trivializes the remaining Virasoro constraint (\ref{V2NATD}). 

Finally, the spinning-string ansatz simplifies the expressions of the conserved charges. Specifically, the energy reduces to
\begin{align}
\label{ENATD} 
E=-\frac{1}{2}\,\tilde{\mu}
\end{align}
and the angular momentum reduces to
\begin{align}
\label{JNATD}
J=\frac{1}{4\pi}\int_{0}^{2\pi}\dif\sigma\frac{r^2}{1+r^2}\,\omega\ .
\end{align}
Note that (\ref{ENATD}), together with (\ref{dispNATD}), states the proportionality between the energy of spinning strings~$E$ and the energy of the mechanical system~$H$. 
Equations (\ref{ENATD}) and~(\ref{JNATD}) are the NATD limit of (\ref{E0}) and (\ref{J0}), respectively. To discuss the connection between spinning strings with $0\leqslant\lambda<1$ 
and their NATD limit, we will use that the NATD limit of the energy can be formally obtained in two steps: first, the replacement~$E\mapsto2kE$; second, the limit $k\rightarrow\infty$.


\section{Spinning strings in the non-Abelian T-dual limit}
\label{sstringsNATD}

In this section, we will analyze the NATD limit of spinning strings. We will address regular spinning strings, and degenerate and nearly degenerate spinning strings separately. 
We will begin with regular spinning strings, which are folded strings, in subsection \ref{ssregularNATD}. We will also evaluate their dispersion relation by numerical means 
and discuss the absence of circular strings afterward. In subsection~\ref{ssndegNATD}, we will turn our attention to point particles, nearly point-like strings, and fast strings. 
We will prove that the dispersion relation of fast strings agrees with the one for GKP strings~\cite{0204051}. In this section, we will consider the NATD limit from the beginning. 
We will comment on the connection of spinning strings in the NATD limit and spinning strings in the regime~$0\leqslant\lambda<1$ when necessary.

To proceed, we will follow the steps of subsection \ref{ssregular}. We start from the \mbox{non-negative} integral of motion $H$ to obtain a first-order ODE for $r$. 
We solve the ODE by a direct integration afterward. If we introduce the moment of inertia~$I$ like in~(\ref{mi}), we obtain
\begin{equation} 
\label{odeNATD} 
r' = \pm \, \omega\sqrt{\frac{(1-2I)(r_{1}^2-r^2)}{1+r^2}} \ ,
\end{equation} 
where we have defined 
\begin{equation}
r_{1}=\sqrt{\frac{2I}{1-2I}} \ .
\end{equation}
The $\pm$ signs in (\ref{odeNATD}) correspond to the positive and negative branches of the square root, respectively. Classical trajectories depend on $\omega$ and $I$. 
We will later trade these parameters by the angular momentum $J$ and the mode number $N$ to write the dispersion relation.

We can obtain (\ref{odeNATD}) in the NATD limit of~(\ref{ode}). To perform the computation, we need two series expansions around $k=\infty$. First, the series of the ellipsoidal coordinate, 
\begin{equation}
\label{zetaNATD}
\zeta=1-\frac{r^2}{4k^2}+\mathrm{O}(k^{-4}) \ ,
\end{equation}
which follows from (\ref{zeta}) and (\ref{rescS}). Second, the series of the parameters
\begin{align}
C&=\frac{1-2I}{k^2}+\mathrm{O}(k^{-3}) \ , \quad\!  \zeta_{1}=1-\frac{I}{2(1-2I)k^2}+\mathrm{O}(k^{-3}) \ , \quad\! \zeta_{2}= 1+\frac{1}{4k^2}+\mathrm{O}(k^{-3}) \ , \nonumber\\
\label{parametersNATD}
I_{1}&=\frac{1}{2k}-\frac{1}{4k^2}+\mathrm{O}(k^{-3})\ , \quad\! I_{2}=\frac{1}{2k}-\frac{1}{4k^2}+\mathrm{O}(k^{-3}) \ .
\end{align}
To obtain the series (\ref{parametersNATD}) from (\ref{parameters}), one needs to introduce the replacement $I\mapsto I/k$ to account for the lack of overall $k$ 
in the definition of the mechanical energy (\ref{H}). 

Classical trajectories must respect the periodic boundary conditions (\ref{bcNATD}). Periodicity is compatible with the ODE (\ref{odeNATD}) if and only if $r^{\prime}$ reaches a turning point, 
that is, a branch point of the square root of (\ref{odeNATD}). The existence of turning points implies both $0\leqslant I\leqslant 1/2$ and $0\leqslant r\leqslant r_{1}$. 
In fact, we can cross-check the condition $0\leqslant I\leqslant 1/2$ by noting that the upper bound $I=1/2$ is the NATD limit of $kI_{1}$. 

\begin{figure}[H]
\centering
{
\includegraphics[width=0.45\textwidth]{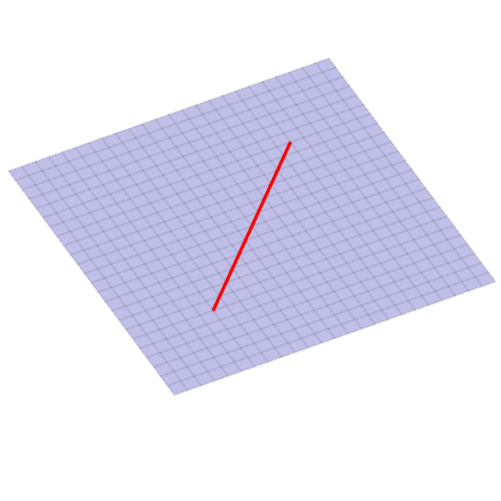}
}
\vskip -1 cm
\caption{Schematic depiction of folded strings in the NATD limit with fixed $\tau$.}
\label{foldedNATDpicture}
\end{figure}

The uniqueness of the range $0\leqslant r\leqslant r_{1}$ reflects the existence of a single regular class of spinning strings. In subsection \ref{ssregularNATD}, 
we will prove that classical trajectories in this regular class realize folded strings. We depict schematically the shape of folded strings in figure \ref{foldedNATDpicture}. 
In subsection \ref{ssndegNATD}, we will prove that the nearly degenerate limit of folded strings gives rise to nearly point-like strings when $J\rightarrow0$ and to fast strings 
when~\mbox{$J\rightarrow\infty$}. The \mbox{spinning-string} ansatz can also realize point particles, which are either instantons or BMN strings. Instantons, which have~\mbox{$J=0$}, 
are connected with folded strings. Moreover, BMN particles, in a sense, bound folded strings.

We first summarize the classification of spinning strings in the NATD limit.

\begin{itemize}
\item{
Folded strings. The inequalities $0<I<1/2$ hold. The range of $r$ is $0\leqslant r\leqslant r_{1}$.
}
\begin{itemize}
\item{
Nearly point-like strings. Folded strings with $I\rightarrow 0$, where $r_{1}\rightarrow0$ and $J\rightarrow0$.
}
\item{
Fast strings. Folded strings with $I\rightarrow 1/2$, where $r_{1}\rightarrow\infty$ and $J\rightarrow\infty$.
}
\end{itemize}
\item{Point particles. The equality $r^{\prime}=0$ holds identically.
\begin{itemize}
\item{
Instantons. Point particles with $r=0$, where $J=0$.
}
\item{
BMN particles. Point particles with $r=\infty$, where $0<J<\infty$.
} 
\end{itemize}
}
\end{itemize}

\subsection{Regular spinning strings}
\label{ssregularNATD} 

In this subsection, we will address regular spinning strings. We will present the case of folded strings, together with their dispersion relation. We will conclude by 
justifying the absence of circular strings.

\subsubsection{Folded strings}

\label{sssfoldedNATD}

Let us first analyze the class of spinning strings with $0<I<1/2$. We need to solve (\ref{odeNATD}). The interval that we must consider is $0\leqslant r\leqslant r_{1}$. 
If we use the invariance of (\ref{odeNATD}) under shifts of~$\sigma$ to set~$r(0)=0$, we must choose the positive branch of~(\ref{odeNATD}). If~$0\leqslant \sigma\leqslant\bar{\sigma}$ 
denotes the interval where $r$ ranges between the turning point of $r^{\prime}$, namely, $r(0)=0$ and~\mbox{$r(\bar{\sigma})=r_{1}$}, direct integration of~(\ref{odeNATD}) leads to
\begin{equation}
\label{solutionNATD}
\begin{split}
\omega\sigma&=\int_{0}^{r}\dif x\sqrt{\frac{1+x^2}{(1-2 I)(r_{1}^2-x^2)}}=\frac{1}{1-2I}\E(\varphi,\sqrt{2I})-\frac{r}{1-2I}\sqrt{\frac{2I-(1-2I)r^2}{1+r^2}} \ ,
\end{split}
\end{equation}
where
\begin{equation}
\varphi=\sin^{-1}\frac{r}{\sqrt{2I(1+r^2)}} \ ,
\end{equation}
and the function $\E(\varphi,m)$ is the incomplete elliptic integral of the second kind, which is analytic and real for~$0<m<1$. The inverse function of~$\E(\varphi,m)$ does not admit a closed form, 
and, hence, $r$ does not have a closed form either.

We could obtain the solution to (\ref{odeNATD}) in any interval other than $0\leqslant\sigma\leqslant\bar{\sigma}$ by alternating the branches of the square root. Since the steps parallel 
those in subsection \ref{sssfolded}, we will omit them. Instead, we will argue that the shape of the string corresponding to (\ref{odeNATD}) is folded. 
(The discussion is analogous to its counterparts of subsections~\ref{sssfolded} and \ref{ssscircular}.)

Let us consider the solution~(\ref{solutionI1}). In the interval $0\leqslant\sigma\leqslant\bar{\sigma}$ , the solution states that the string is stretched from the center 
of $\widetilde{\mathrm{R}}^{2}$ at $r(0)=0$ along the radial ray specified by~$\gamma=\omega\tau$ until $r(\bar{\sigma})=r_{1}$. 
The negative branch of~$r^{\prime}$ begins at the turning point~\mbox{$r(\bar{\sigma})=r_{1}$}. Along~\mbox{$\bar{\sigma}\leqslant\sigma\leqslant2\bar{\sigma}$}, the radius $r$ ranges through the same path as before
but backward. The value~\mbox{$r(2\bar{\sigma})=0$} is not a turning point of~$r'$. To continue~$r$, we need to perform the shift $\gamma\mapsto\gamma+\pi$, 
because the embedding coordinates~(\ref{embedding}) and the rescaling~(\ref{rescS}) imply that $\gamma\mapsto\gamma+\pi$ is equivalent to $r\mapsto-r$. 
The shift, in turn, effectively translates into the change to the positive branch of $r^{\prime}$. Once we change the branch of $r^{\prime}$, 
the path of~$r$ along~\mbox{$2\bar{\sigma}\leqslant\sigma\leqslant 4\bar{\sigma}$} coincides with the path along $0\leqslant\sigma\leqslant 2\bar{\sigma}$. 
The string reaches the initial point at~$r(4\bar{\sigma})=0$. The periodic boundary conditions~(\ref{bcNATD}) imply that~$r$ must cover the path over~$0\leqslant\sigma\leqslant4\bar{\sigma}$ 
an integer number of times, and, thus, $r(\sigma)=r(2\pi-\sigma)$. The discussion implies that the string is folded and centered at $r=0$ as depicted in figure~\ref{foldedNATDpicture}.

\begin{figure}[H]
\centering
\subfloat[]{\includegraphics[width=0.45\textwidth]{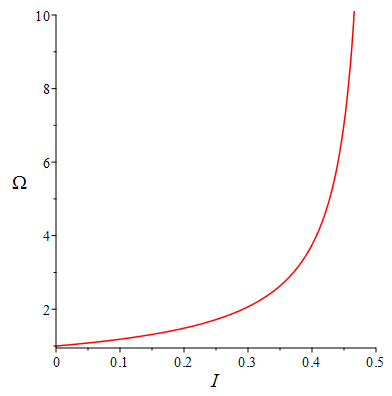}\label{omegaNATD}}
\subfloat[]{\includegraphics[width=0.45\textwidth]{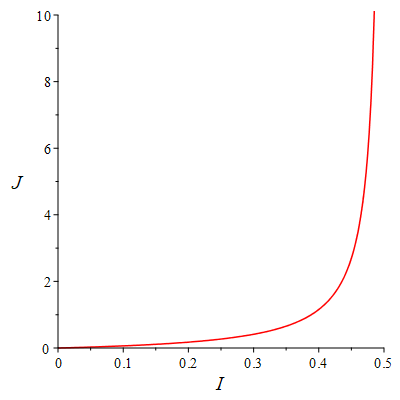}\label{JNATDfig}}
\caption{The quantities $\Omega=\omega/N$ and $J/N$ of folded strings through $0<I<1/2$ in the NATD limit.}
\end{figure}

From the previous discussion we conclude that $r$ must go through $0\leqslant r\leqslant r_{1}$ a multiple of four number of times, and, thus, 
a quantization condition formally identical to (\ref{qc}) must hold. In this case,~$N$ again counts the number of times that the string folds, 
like for the classical solutions of subsection \ref{sssfolded}. Explicitly, it follows from (\ref{qc}) and (\ref{solutionNATD}) that
\begin{equation}
\label{NNATD}
\Omega\equiv\frac{\omega}{N}=\frac{2}{\pi(1-2I)}\E(\sqrt{2I}) \ , 
\end{equation}
where $\E(m)$ is the complete elliptic integral of the second kind, which is analytic and real for $0<m<1$. We have plotted~(\ref{NNATD}) in figure \ref{omegaNATD}. 
The right-hand side of~(\ref{NNATD}) is analytic, positive, and strictly increasing. At the left endpoint, we have
\begin{equation}
\label{NleftNATD}
\Omega\rightarrow1 \ , \quad I\rightarrow0 \ .
\end{equation}
We must exclude (\ref{NleftNATD}) because $I=0$ implies $r=0$, where $\gamma$ is not defined. At the right endpoint, we have
\begin{equation}
\label{NrightNATD}
\Omega\rightarrow\infty \ , \quad I\rightarrow\frac{1}{2} \ .
\end{equation}
Since $\E(1)=1$, the divergence of follows from the simple pole of (\ref{NNATD}). 

We should note that (\ref{NNATD}) is the NATD limit of (\ref{NI}). To prove the claim we should perform the series expansion of (\ref{NI}) around $k=\infty$. 
(Recall that we must introduce the replacement~$I\mapsto I/k$ to apply the NATD limit to (\ref{NI}).) The series is straightforward but cumbersome, 
and we will not present it here.  Nonetheless, it is worth making two observations regarding the computation. First, $n=\mathrm{O}(k^{-2})$ in (\ref{argumentsI}), 
which implies that we must perform the series of $\Pi(n,m)$ around $n=0$. Second, terms that would appear at $\mathrm{O}(k^2)$ and $\mathrm{O}(k)$ in~(\ref{NI}) turn out to vanish, whereby just (\ref{NNATD}) remains.

Moreover, the angular momentum $J$ given by (\ref{JNATD}) determines the first integral $I$ implicitly. To integrate (\ref{JNATD}), 
we will follow the steps of subsection \ref{ssregular} and split~\mbox{$0\leqslant\sigma\leqslant2\pi$} into $4N$ subintervals of equal length. 
Since the contribution to the interval of each subinterval is the same, we can rephrase the integral as $4N$ times the integral over $0\leqslant\sigma\leqslant\bar{\sigma}$. If we use (\ref{odeNATD}), 
we find that (\ref{JNATD}) eventually becomes 
\begin{equation}
\begin{split}
\label{J1NATD}
\frac{J}{N}=\frac{1}{\pi}\int_{0}^{r_{1}}\dif x\frac{x^2}{\sqrt{{(1-2 I)(1+x^2)(r_{1}^2-x^2)}}}
&=\frac{1}{\pi}\left(\frac{1}{1-2I}\E(\sqrt{2I})-\K(\sqrt{2I})\right) \ .
\end{split}
\end{equation}
We have plotted~(\ref{J1NATD}) in figure \ref{JNATDfig}. The right-hand side of~(\ref{J1NATD}) is analytic and real. Figure \ref{JNATDfig} shows that it is also positive and strictly increasing. 
Therefore, $J$ and $N$ have the same sign, and $J$ is positive. On the left endpoint of $0<I<1/2$, we have
\begin{equation}
\label{JleftNATD}
\frac{J}{N}\rightarrow 0 \ , \quad I\rightarrow 0 \ .
\end{equation}
The vanishing is consistent with the fact that $\gamma$ is not defined at $r=0$. At the right endpoint, we have
\begin{equation}
\label{JrightNATD}
\frac{J}{N}\rightarrow \infty \ , \quad I\rightarrow \frac{1}{2} \ .
\end{equation}

We will close the discussion by noting that (\ref{J1NATD}) is the NATD limit of (\ref{JI}). The application of the NATD limit is straightforward but cumbersome again, 
and we will not present it here. We will just note that, similarly to what we have commented on before, terms at~\mbox{$\mathrm{O}(k^2)$} and~$\mathrm{O}(k)$ in the series of~(\ref{JI}) around $k=\infty$ cancel, leaving the finite remnant~(\ref{J1NATD}).

\subsubsection{The dispersion relation}

\label{sssdispNATD}

Let us now address the dispersion relation of folded strings. Before we turn to the computation itself, we will complete our discussion in subsection \ref{sssdisp} 
by arguing that the NATD limit can be deemed to be a semiclassical limit. In what follows we will restore $\kappa$ to make the discussion clearer.

The NATD limit involves the limit $k\rightarrow\infty$ and the rescaling of the target-space coordinates. Moreover, $k\rightarrow\infty$ is the semiclassical limit of the path integral that gives rise to spinning strings. 
If~$0\leqslant\lambda<1$, formula~(\ref{lambda}) states that the limit $\kappa\rightarrow\infty$ must supplement 
the semiclassical limit $k\rightarrow\infty$.~\footnote{Reference \cite{1704.05437} made an analogous observation in the $\lambda$-deformation of $\mathrm{AdS}_{5}\times\mathrm{S}^{5}$.} 
On the other hand,~$k\rightarrow\infty$ implies~\mbox{$\lambda=1$} for every finite~$\kappa$. 
Therefore, whenever $\kappa$ is finite, the NATD limit is identifiable with the semiclassical limit of the type IIB supergravity background that embeds the~$\lambda$-deformation of the $\SU$ WZW model.
This observation applies everywhere but at the undeformed limit because $\kappa\rightarrow\infty$ is a necessary condition for~$\lambda=0$ to hold. In this sense, 
spinning strings on $\widetilde{\mathrm{R}}^{3}$ are the accurate realization of the solitons of the conformal field theory of the superstring worldsheet that we have sketched in subsection \ref{sssdisp}.

Let us now compute the dispersion relation of folded strings. To compute the dispersion relation, we must follow the steps of subsection~\ref{sssdisp}. 
First, we must express $\omega$ and~$I$ in terms of $J$ and $N$ through equations~(\ref{NNATD}) and~(\ref{J1NATD}). Since the correspondence 
between~\mbox{$0<I<1/2$} and $0<J/N<\infty$ is one-to-one and onto, as figure~\ref{JNATDfig} shows,~(\ref{J1NATD}) determines~\mbox{$I=I(J/N)$} unambiguously. 
Figure \ref{omegaNATD} implies that~(\ref{NNATD}) fixes~$\omega=N\Omega(J/N)$ unambiguously as well. Finally, the Virasoro constraint (\ref{dispNATD}) together 
with formula (\ref{ENATD}) enables us to write $E$ in terms of $N$ and $J/N$ through $\Omega$ and $I$. 

\begin{figure}[h]
\centering
\includegraphics[width=0.45\textwidth]{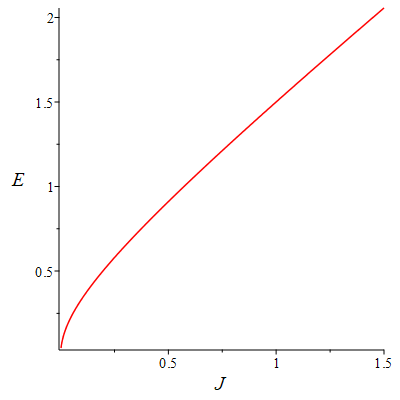}
\caption{The dispersion relation of spinning strings with $N=1$ in the NATD limit.}
\label{dispNATDfig}
\end{figure}

The formula of the dispersion relation that we should consider is
\begin{equation}
\label{dispNATD2}
E=N\Omega(J/N)\sqrt{\frac{I(J/N)}{2}} \ ,
\end{equation}
where we have chosen $E$ to be positive in relating the energy and the Virasoro constraints. We cannot compute (\ref{dispNATD2}) analytically in general, but just in special limits. 
We defer the analytical computation of the dispersion relation until subsection \ref{ssndegNATD}. In this subsection, we will evaluate instead (\ref{dispNATD2}) numerically. 
We have plotted the results in the case with~$N=1$ in figure~\ref{dispNATDfig}. There is no loss of generality in the choice of $N=1$ because the mode number just appears in the ratio $E/N$ and $J/N$.

Figure \ref{dispNATDfig} resembles the dispersion relation of folded strings on $\mathrm{AdS}_{3}$ (see, for instance, figure~6 of \cite{1311.5800}). 
We will bear the resemblance out in subsection~\ref{sssfastNATD}, where we will prove that fast strings share the dispersion relation with GKP strings at leading order.

We will conclude the discussion on (\ref{dispNATD2}) with a comment on the connection between the dispersion relation in the regime $0\leqslant\lambda<1$ and in the NATD limit. 
We have already remarked that we can apply the NATD limit to the energy by means of the replacement~\mbox{$E\mapsto 2kE$} and the subsequent limit $k\rightarrow\infty$. 
However, it is not apparent that the substitution provides the correct result from the numerical point of view. We will prove in the next subsection that the NATD limit
of the dispersion relation of point particles, and the leading order of the dispersion relation of nearly point-like strings and fast strings, 
follow from these steps. In fact, these two steps also help to clarify the lack of circular strings, as we will now discuss.

\subsubsection{Absence of circular strings}
\label{ssscircularNATD} 

Unlike folded strings, circular strings are absent in the NATD limit.  We can explain the situation from two complementary points of view. 
First, we can apply the NATD limit from the outset. In this case, we can argue that circular strings are forbidden since they would surround $\widetilde{\mathrm{R}}^{2}$ 
along a non-compact direction. More precisely, if we solved (\ref{lambdaNATD}) with~$1/2<I<\infty$, we could not impose the periodicity of $r$ that we have written in~(\ref{bcNATD}), 
and the angular momentum would diverge.~\footnote{One could draw an analogy with spinning strings in $\mathrm{AdS}_{5}\times\mathrm{S}^{5}$. If spinning strings 
with $0<I< 1/2$ correspond to GKP strings (as we will argue), spinning strings with $1/2<I<\infty$ should correspond to minimal surfaces dual to Wilson loops with a null cusp \cite{0210115}. 
The limitation of the analogy is apparent however. First, there is no reason for minimal surfaces to exist in $\widetilde{\mathrm{R}}^{3}$. Second, no definition of boundary of $\widetilde{\mathrm{R}}^{3}$ has not been provided yet.} 
Second, we can argue the absence of circular strings by applying the NATD limit to the regime $0\leqslant\lambda<1$. The NATD limit of the quantities 
$\Omega$ and $J$ of folded strings is finite and given by~(\ref{NNATD}) and~(\ref{J1NATD}). On the contrary, the NATD limit of $\Omega$ and $J$ of circular strings diverges. If we performed the series of $\Omega$ in~(\ref{NII}) around~$k=\infty$, 
we would obtain that terms at $\mathrm{O}(k)$ do not vanish. If we performed the series of $J$ in~(\ref{JII1}) around~$k=\infty$, we obtain that terms at $\mathrm{O}(k)$ 
do not vanish either. Since $E$ is proportional to $\Omega$ through (\ref{disp2}), the divergence~$\Omega\rightarrow\infty$ implies the divergence~\mbox{$E\rightarrow\infty$}. 
Therefore, circular strings lack in the NATD limit because they cannot be reached from an energetic point of view. The divergence of the NATD limit of~$E$ 
for circular strings is also consistent with figures \ref{disp1} and \ref{disp2plot}, in the sense that the gap between folded and circular strings becomes insurmountable in the NATD limit. 
Additional support comes from the NATD limit of the energy of slow strings, because, if we had introduced $E\mapsto 2kE$ in the dispersion relation~(\ref{Eslow}) and had applied the limit~\mbox{$k\rightarrow\infty$} afterward, we would have obtained the scaling $E\sim k$. 

\subsection{Point particles and nearly degenerate spinning strings}
\label{ssndegNATD} 

In this subsection, we will consider degenerate and nearly degenerate spinning strings, for which we will compute the dispersion relation at leading order in $J$. 
We will first address the case of point particles, to turn to nearly point-like strings later on.  We will conclude the analysis with the case of fast strings, which we will prove they have 
the dispersion relation of GKP strings.

\subsubsection{Point particles}
\label{ssspointNATD} 

We will start the discussion with point particles, which are trivially periodic spinning strings with $N=0$. Since $r^{\prime}=0$, we must check that the ansatz (\ref{ansatzNATD}) consistently truncates the equations of motion of the action (\ref{SNATD}) again. The only consistent choices are $r=0$ and $r=\infty$.

If $r=0$, the angle $\gamma$ is not defined. The Virasoro constraints~(\ref{V1NATD}), the energy~(\ref{ENATD}), and the angular momentum (\ref{JNATD}) imply $E=J=0$. The point particle is localized at the center of $\widetilde{\mathrm{R}}^{3}$ and $\tilde{t}=0$,
and we can argue that the point particle is an instanton for this reason. Moreover, the radius of the orbit is infinite when~\mbox{$r=\infty$}. The conserved charges $E$ and $J$ are still finite nonetheless, and
\begin{equation}
\label{EBMNNATD}
E = J \ ,
\end{equation}
which is the usual dispersion relation of BMN particles. 

The dispersion relation~(\ref{EBMNNATD}) is the NATD limit of the dispersion relation of BMN particles~(\ref{EBMN}). 
Therefore, BMN particles in the NATD limit are point particles that orbit the center of~$\widetilde{\mathrm{R}}^{3}$ along a circumference at infinity and, in a sense, bound folded strings.

\subsubsection{Nearly point-like strings}

\label{sssnpointNATD}

Nearly point-like strings are folded strings with either small $J$ or large~$N$. Formula~(\ref{JleftNATD}) states that $J/N\rightarrow 0$ when $I\rightarrow0$. 
Since $I\rightarrow0$ implies $r_{1}\rightarrow0$, we deduce that~$0\leqslant r\leqslant r_{1}$ tends to collapse, and, hence, strings are ``nearly point-like". 
If we perform the series expansion of (\ref{solutionNATD}) around $r=0$ and $I=0$ with  $r/\sqrt{I}$ fixed, we obtain
\begin{equation}
\label{rnpointNATD}
r=\sqrt{2I}\abs{\sin(\omega\sigma)}+\mathrm{O}(I) \ ,
\end{equation}
where we have used $\E(\varphi,0)=\varphi $.
The classical solution (\ref{rnpointNATD}) is the NATD limit of (\ref{thetanpoint}). To apply the NATD limit, we must restore~$\zeta=\cos^2\theta$ in (\ref{thetanpoint}), 
replace $I\mapsto I/k$ therein, and shift $\sigma\mapsto\sigma+\pi/2$. It is worth noting that (\ref{rnpointNATD}), just as (\ref{thetanpoint}), resembles the radial coordinate 
$r$ of folded strings in flat space, which is indeed consistent with the expected suppression of the effects of the target-space curvature for almost point-like strings.

We can support the analogy between nearly point-like strings and strings in flat space by computing the dispersion relation at leading order. If we perform the series 
of (\ref{NNATD}) and (\ref{J1NATD}) around $I=0$, we obtain
\begin{equation}
\Omega = 1 + \mathrm{O}(I) \ , \quad \frac{J}{N}=\frac{1}{2}I+\mathrm{O}(I^2) \ ,
\end{equation}
which, by means of (\ref{ENATD}), imply the Regge dispersion relation
\begin{equation}
\label{EnpointNATD}
E = \sqrt{NJ} +\mathrm{O}(J) \ .
\end{equation}
The dispersion relation (\ref{EnpointNATD}) is the NATD limit of~(\ref{Enpoint}).

\subsubsection{Fast strings}
\label{sssfastNATD}

We will now address fast strings, which are folded strings with large angular momentum. Formula (\ref{JrightNATD}) states that~$J\rightarrow\infty$ when \mbox{$I\rightarrow 1/2$}. 
Since~(\ref{NrightNATD}) implies $\omega\rightarrow\infty$, the rotations of these strings are referred to as ``fast". Let us consider first the dispersion relation. We will start 
with the computation of $I=I(J/N)$ as a series around $J=\infty$. The limit $I\rightarrow1/2$ 
implies $m\rightarrow1$ for $\K(m)$ and $\E(m)$ in the expression of the total angular momentum (\ref{J1NATD}). If we use the asymptotic series
\begin{align}
\K(m)&\simeq-\frac{1}{2}\log(1-m^2)+\log4 \ , \quad m\rightarrow 1 \ , \\
\E(m)&\simeq 1+\frac{1}{4}[\,2\log4-1-\log(1-m^2)\,](1-m^2) \ , \quad m\rightarrow 1 \ ,
\end{align}
we obtain
\begin{equation}
\label{IGKP}
\frac{4}{1-2I}\simeq W(x) \ , \quad J\rightarrow\infty  \ ,
\end{equation}
where $x\equiv 64\exp(4\pi J / N)$ and $W$ is the Lambert W function in the principal branch \mbox{$W:[-1/\e,\infty)\rightarrow[-1,\infty)$}. 
(Recall that the function $W$ is real and monotonically increasing.) If we use the asymptotic series
\begin{equation}
W(x)\simeq\log x-\log(\log x) \ , \quad x\rightarrow\infty \ ,
\end{equation}
we obtain from (\ref{IGKP}) and (\ref{NNATD}) 
\begin{equation}
\label{IGKP2}
I\simeq\frac{1}{2} \ , \quad J\rightarrow\infty \ ,
\end{equation}
and 
\begin{equation}
\label{omegaGKP}
\Omega\simeq\frac{2J}{N}+\frac{1}{\pi}\log\frac{16\pi J}{N} \ , \quad J\rightarrow\infty \ ,
\end{equation}
respectively. If we introduce (\ref{IGKP2}) and (\ref{omegaGKP}) into (\ref{dispNATD2}), we conclude that 
\begin{equation}
\label{EGKP}
E\simeq J+\frac{N}{2\pi}\log\left(\frac{16\pi J}{N}\right) \ , \quad J \rightarrow\infty \ .
\end{equation}
The dispersion relation (\ref{EGKP}) matches exactly the dispersion relation of GKP strings under the formal identifications of the 't Hooft coupling~$\sqrt{\lambda}=N/2$ and the Lorentzian spin $S=J$ (see, 
for instance, equation (4.5) of reference \cite{1012.3986}). 
Relation (\ref{EGKP}) should be the dispersion relation of fast strings with $0\leqslant\lambda<1$ in the NATD limit. 
Note that even though we can obtain the linear term in (\ref{EGKP}) from~(\ref{Efast}), we cannot obtain the logarithmic term because we have not computed analytically subleading terms in (\ref{Efast}).

We do not write the explicit solution because it does not admit a closed form. Nonetheless, we can describe the change of folded strings in the limit of fast rotations. 
The limit~\mbox{$I\rightarrow 1/2$} pulls the endpoints of the string at radius $r=r_{1}$ toward $r=\infty$, which enlarges the size of the string. Besides, the limit $I\rightarrow 1/2$ also increases 
the angular speed because $\omega\rightarrow\infty$, as we have already mentioned. Therefore, the longer the string is, the faster the string rotates. When the string is infinitely stretched, periodicity cannot hold anymore, 
rotations are infinitely fast, and the string disintegrates.


\section{Conclusions}
\label{sconclusions}

In this article, we have analyzed spinning strings that propagate through the target space of the $\lambda$-deformation of the $\mathrm{SU}(2)$ WZW model. 
We have assumed the existence of the embedding into a special type IIB supergravity background to define the energy and remove the NSNS flux.
We have considered the regime $0\leqslant\lambda<1$ and the NATD limit separately. If~$0<\lambda<1$, we found that the classification of spinning strings 
parallels the situation at~\mbox{$\lambda=0$}, where regular spinning strings are either folded or circular, and nearly degenerate spinning strings are either nearly point-like, 
fast, or slow strings. We have found that the $\lambda$-deformation increases the overall energy of spinning strings and broadens the gap between the energy of folded and circular strings. 
In the NATD limit, where~$\lambda=1$, we have proven that the energy of circular strings diverges and that just folded strings remain valid solutions.  
If these folded strings are large and rotate fast, we have obtained that the dispersion relation, namely~(\ref{EGKP}) matches the dispersion relation of GKP strings. 
On the whole, our analysis of spinning strings suggests that the semiclassical approach can shed new light on the \mbox{$\lambda$-deformation}, and, 
in particular, on the radical difference between the regime~\mbox{$0\leqslant\lambda<1$} and the NATD limit. Our work hints at multiple problems, some of which we comment on below.

First and foremost, to be complete, our analysis demands the construction of the embedding into type IIB supergravity of the $\lambda$-deformation of the $\mathrm{SU}(2)$ WZW model.
The consistency of the spinning-string ansatz strongly relied on the removal of the NSNS flux by S-duality, and the dispersion relation resorted to the existence of a globally defined time-like Killing vector. 
To proceed, one may use our assumption on the RR three-form flux, namely, the absence of components along $\mathrm{S}_{\lambda}^{3}$,  the constant dilaton, the existence of time-like Killing vector, and the preservation of the $\SU_{R}$ symmetry group to make an educated guess for the ansatz.
Furthermore, it would be highly desirable that the embedding into type IIB supergravity involved an AdS truncation or a deformation thereof. This property would bring spinning strings here closer to holography and the AdS/CFT correspondence.

Moreover, in addition to spinning strings, pulsating strings constituted an important class of classical solutions in the AdS/CFT correspondence. 
First proposed by \cite{0209047}, pulsating strings can be defined as the classical solutions implied by the spinning-string ansatz under the interchange 
of $\tau$ and $\sigma$~\cite{0311004}. Pulsating strings and spinning strings are sharply different nonetheless. The specification of Noether charges and winding numbers 
of pulsating strings is straightforward due to the interchange between $\tau$ and $\sigma$. On the other hand, periodic boundary conditions in $\sigma$ are fulfilled automatically, 
and, thus, the introduction of an adiabatic invariant~$M$ becomes necessary~\cite{9410219}. If~$M$ is semiclassical and large, the energy~$E$ of pulsating strings admits 
a series expansion in the adiabatic invariant that mimics the series of the energy~$E$ of spinning strings in~$J$~\cite{0209047,0310188,0406189}. 
In the embedding of the $\lambda$-deformed~$\SU$ WZW model into type IIB supergravity we have considered, pulsating strings could illuminate various points.  First, they would make more precise the extent to which effects 
that we have ascribed to the $\lambda$-deformation, such as the increment of the energy of spinning strings and the enlargement of the energetic gap between regular classes, may extend  
to other components of the semiclassical spectrum. Second, pulsating strings may clarify if other classical solutions apart from circular spinning strings can become energetically forbidden in the NATD limit. 
The matching between~(\ref{EGKP}) and the energy of GKP strings also points toward another direction. One may endeavor to map the energy~$E$ of the NATD limit of pulsating strings in the limit of semiclassical 
and large~$M$ into~$E$ of pulsating strings through~$\mathrm{AdS}_{3}$. If this mapping exists, one would obtain favorable support for the existence of a general mapping 
between the NATD of the~$\SU$ PCM in the S-dual framework and the~$\SL$ PCM in the sector of semiclassical and large quantum numbers. 

Finally, it could be worth exploring spinning strings in the $\lambda$-deformation of other nonlinear $\sigma$-models.
An example of $\lambda$-deformation that is closely related to the framework of our work has been considered in \cite{1707.05149}. The $\lambda$-deformation in this case couples 
two WZW models based on the Lie group~$\mathrm{G}$ at levels~$k_{1}$ and~$k_{2}$. The exact $\lambda$-deformation turns out to coincide with a perturbative \mbox{current-current} 
deformation that couples the two models and preserves a subgroup~$\mathrm{G}\times\mathrm{G}$ of the full isometry group of the joint pair of models. 
When the deformation parameter reaches the value~$\lambda=\sqrt{k_{1}/k_{2}}$ (under the assumption $k_{1}<k_{2}$), the $\lambda$-deformation corresponds 
to two non-coupled WZW models at levels~$k_{1}$ and $k_{2}-k_{1}$.  The $\lambda$-deformation of \cite{1707.05149} poses new challenges for the realization of spinning strings 
by means of mechanical systems since it interpolates between pairs of non-coupled WZW models. More specifically, the use of the spinning-string ansatz of \cite{0311004} 
would reduce the equations of motion of the model of \cite{1707.05149} to the ODEs of the $\lambda$-deformation of a mechanical system based on $\mathrm{G}\times\mathrm{G}$. 
If the $\lambda$-deformation reaches one of the two endpoints, that is, if either~$\lambda=0$ or $\lambda=\sqrt{k_{1}/k_{2}}$, the $\lambda$-deformation of the complete mechanical system 
must reduce to two non-coupled copies of the system of \cite{1502.05203}. Since the maximally deformed point of the system at $\lambda=0$ is the undeformed point of the system 
at $\lambda=\sqrt{k_{1}/k_{2}}$, and vice versa, the $\lambda$-deformation of \cite{1707.05149} poses the problem of the integrable interpolation between two realizations of the same mechanical system. 
A promising starting point to tackle the problem could probably come from the use of canonical transformations.

\subsectionfont{\centering}

\subsection*{Acknowledgements}

\noindent The work of R. H. and R. R. has been supported through the grant PGC2018-095382-B-I00, and by Universidad Complutense de Madrid 
and Banco de Santander through the grant GR3/14-A 910770. R. R. has been supported by Universidad Complutense de Madrid and Banco de Santander through the contract CT42/18-CT43/18. 
R. R. thanks J. M. Nieto Garc{\'i}a for valuable discussions. 
The authors thank the referee for pointing out an error in the original version of the manuscript.
R. R. is grateful to the Department of Physics 
of the National and Kapodistrian University of Athens 
for kind hospitality during the development of this work. The work of K. S. was supported by the Hellenic Foundation for Research and Innovation (H.F.R.I.) 
under the ``First Call for H.F.R.I. Research Projects to support Faculty members and Researchers and the procurement of high-cost research equipment grant" (MIS 1857, Project Number: 16519).

\bibliographystyle{ieeetr}

\bibliography{SpinningLambda}{}

\end{document}